\documentclass[a4paper,11pt]{article}
\pdfoutput=1
\usepackage{jheppub}
\usepackage{amsmath}
\usepackage{amssymb}
\usepackage{dcolumn}	
\usepackage{bm}			
\usepackage{bbm} 		
\usepackage{multirow}
\usepackage{blkarray}
\usepackage{slashed}
\usepackage{bbding}
\usepackage{pifont}
\usepackage[usenames,dvipsnames]{xcolor}
\definecolor{hgreen}{rgb}{0,.3,0}
\definecolor{hred}{rgb}{.3,0,0}
\definecolor{hblue}{rgb}{0,0,.3}
\definecolor{LightGray}{gray}{0.95}
\makeatletter
\def\endfmffile{%
	\fmfcmd{\p@rcent\space the end.^^J%
		end.^^J%
		endinput;}%
	\if@fmfio
	\immediate\closeout\@outfmf
	\fi
	\ifnum\pdfshellescape>\z@
	\immediate\write18{mpost \thefmffile}%
	\fi}
\makeatother

\usepackage{pbox}
\usepackage{caption}
\usepackage{subcaption}
\usepackage{xcolor}
\usepackage[utf8]{inputenc}

\graphicspath{{./Figure/}}

\newcommand{\pid}{\hat{\pi}}

\newcommand{\bomega}{\bm{\omega}}
\newcommand{\bmass}{\bm{M}}
\newcommand{\byuk}{\bm{Y}}
\newcommand{\byukt}{\widetilde{\bm{Y}}}

\newcommand{\bmpsi}{\bm{m}_\psi}
\begin{document}


\title{Phenomenology of Electroweak Portal Dark Showers: High Energy Direct Probes and Low Energy Complementarity}

\author[a]{Hsin-Chia Cheng,}
\affiliation[a]{Center for Quantum Mathematics and Physics (QMAP), Department of Physics,\\ University of California, Davis, USA}
\author[b,c]{Xu-Hui Jiang,}
\author[d]{Lingfeng Li}
\affiliation[b]{{Center for Future High Energy Physics, Institute of High Energy Physics, Chinese Academy of Sciences, Beijing, China}}
\affiliation[c]{{China Center of Advanced Science and Technology, Beijing, China}}
\affiliation[d]{Department of Physics and Brown Theoretical Physics Center, Brown University,\\ Providence, USA}
\date{\today}
\emailAdd{cheng@physics.ucdavis.edu, jiangxh@ihep.ac.cn, lingfeng\_li@brown.edu}

\abstract{We investigate the phenomenology of a dark QCD sector interacting with the Standard Model (SM) via the electroweak (EW) portals. The portal interactions allow SM bosons, such as $Z$ and $h$, or additional bosons that mix with them, to decay into dark quarks, producing dark showers. The light dark mesons are expected to be long-lived particles (LLPs), as their decays back to the SM states through the EW-portal interactions typically have macroscopic decay lengths. We focus on dark shower events initiated by various bosons at the Large Hadron Collider (LHC). The most prominent signal is the displaced decay of GeV-scale dark pions as LLPs. Current limits on dark shower signals at LHC detectors are recast from public data to provide simplified limits insensitive to UV physics details. Future limits in the high-luminosity phase and proposed auxiliary detectors are also projected. Additionally, we study the flavor-changing neutral current (FCNC) $B$ decays into dark pions, obtaining both current and projected constraints at the LHC and other facilities. These constraints can be combined for specific models, which are illustrated in two EW-portal benchmarks: one with the heavy doublet fermion mediation and another with the $Z^\prime$ mediator including a mass mixing. The collider reach shows significant potential to probe the parameter space unconstrained by EW precision tests, highlighting the necessity of dedicated LLP search strategies and facilities.}

\maketitle


\section{Introduction}
\label{sec:Intro}

A dark sector containing a quantum chromodynamics (QCD)-like structure which interacts very weakly with the Standard Model (SM) appears in many new physics scenarios, motivated by important questions of nature, e.g., the neutral naturalness~\cite{Chacko:2005pe,Burdman:2006tz,Cai:2008au,Cheng:2018gvu,Cohen:2018mgv} or cosmological relaxation~\cite{Graham:2015cka} solutions to the hierarchy problem. It also provides possible candidates of the dark matter (DM) with dark hadrons if they are cosmologically stable. Experimentally, if the dark quarks or gluons are produced at energies much higher than the confining scale, such as in a collider environment, they are expected to generate dark showers and end up with multiple dark hadrons. Depending on whether these dark hadrons decay back to SM particles and their lifetimes, they can give rise to a variety of exotic collider signals, including displaced decays of long-lived particles (LLPs)~\cite{Strassler:2006im,Strassler:2006qa,Han:2007ae,Alimena:2019zri}, emerging jets~\cite{Schwaller:2015gea}, semi-visible jets~\cite{Cohen:2015toa}, etc. Therefore, these models are also of great experimental interest to explore~\footnote{See also Ref.~\cite{Park:2017rfb} for related phenomenology.}.

The collider phenomenology of the dark QCD depends on the portal interactions between the SM and the dark sectors, which govern both dark particle production and dark hadron decays. If the dark sector and the SM only couple through some very heavy ($\gtrsim$ TeV) states, at low energies the mediators can be integrated out to induce higher-dimensional effective interactions between singlet SM operators and dark operators. Without additional light mediators, it is expected that a significant fraction (if not all) of the light dark hadrons is long-lived or stable. If the SM singlet operator that couples to the dark sector always involves two or more SM particles, the production of the dark particles is enhanced at energies near the mediator scale.  At the Large Hadron Collider (LHC), the dark particles are then typically produced with high energies, including the possibilities of associated energetic visible particles or jets, which could help trigger the events.  The dark showers induced by the dark quarks or gluons can give rise to a variety of experimental signatures, depending on the dark QCD dynamics and the portal interactions. They produce semi-visible jets if only a fraction of the dark hadrons give visible decays. If the dark hadrons have displaced decays within the detector, the corresponding jets will appear as emerging jets. Searches for prompt semi-visible jets have put strong bounds on the scales that suppress higher dimensional operators or the heavy mediator masses~\cite{Cohen:2017pzm,CMS:2021dzg,ATLAS:2023swa}. For the $s$-channel $Z'$ mediator with a coupling 0.25 to SM quarks, the $Z'$ mass can be constrained for the invisible fraction $0.01 < r_{\rm inv} < 0.78$. It is excluded from 1.5~TeV up to 5.1~TeV for $r_{\rm inv}$ around 0.3 and between 2.1~TeV and 3.3~TeV for $r_{\rm inv}\sim 0.7$~\cite{CMS:2021dzg}. For the $t$-channel model where the mediator couples to an SM quark and a dark quark, the mediator mass is constrained to be above 2.4--2.7 TeV for a coupling $\lambda=1$ and $0.1 < r_{\rm inv} < 0.9$~\cite{ATLAS:2023swa}. The existing emerging jet search is based on a specific model where a heavy mediator particle decays to an SM jet and an emerging jet~\cite{CMS:2018bvr,CMS:2024rea}. The heavy mediator mass is excluded up to 1.95~TeV for the dark hadron decay length less than 100 mm, and 1.2~TeV for decay length $\sim 1$ m. If all dark hadrons have long lifetimes and are stable at the collider scale, the dark showers will be invisible, and the collider bounds come from mono-X searches just as the usual DM searches. For an $s$-channel vector mediator, the monojet searches have put a $\sim 2$ TeV bound on the $Z'$ mass with a coupling 0.25 to SM quarks~\cite{CMS:2021far,ATLAS:2021kxv}.

\begin{figure}
\centering
\includegraphics[width=0.95\textwidth]{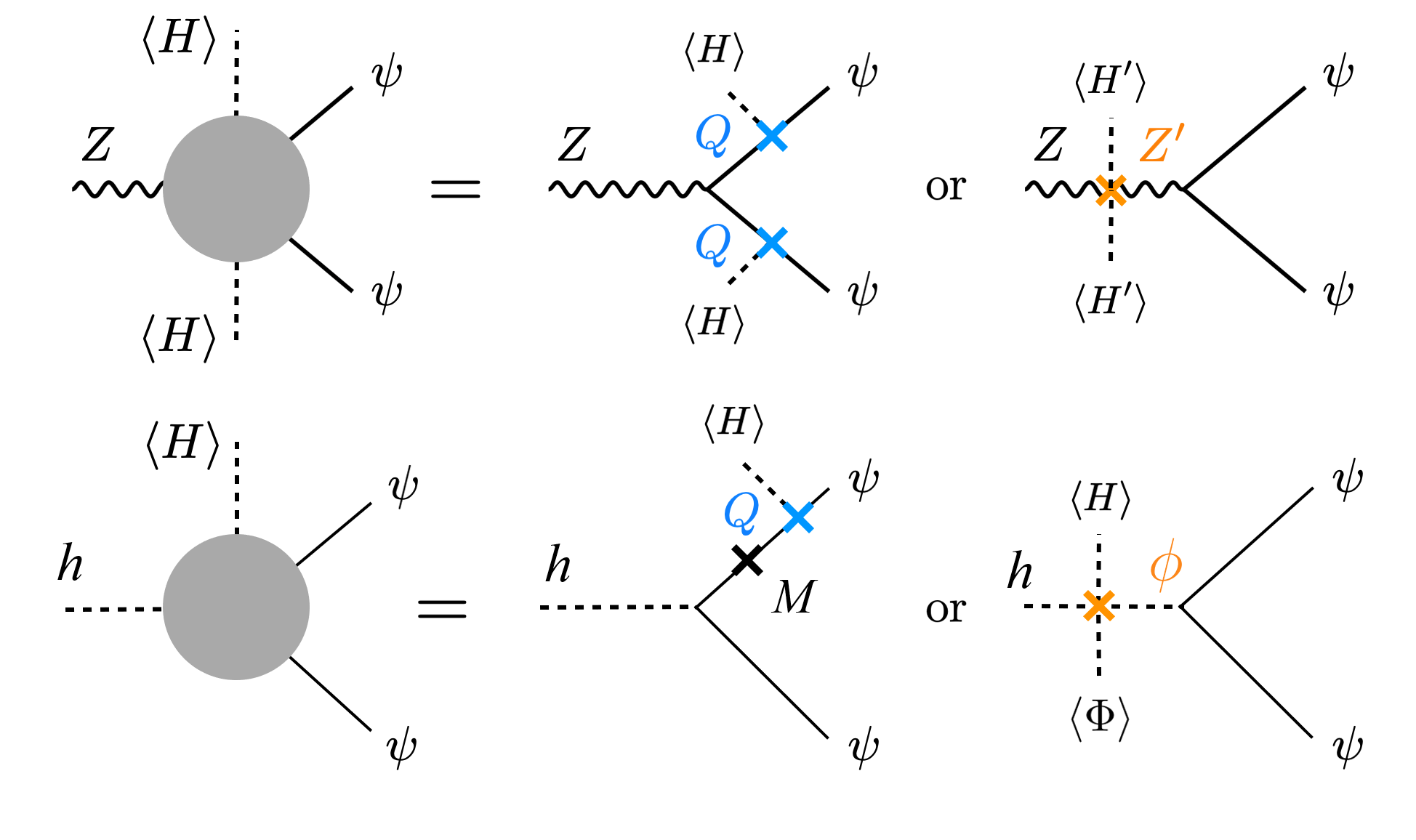}
\caption{The EW-portal operators in Eq.~\eqref{eq:Zh_operators} and the two underlying models. \textbf{MIDDLE:} the heavy fermion mediation introduced in Ref.~\cite{Cheng:2021kjg}. In this case, the heavy dark quark $Q$, an SM EW doublet, obtains a mixing with the light dark quark $\psi$ via the Yukawa interaction with the SM Higgs field. \textbf{RIGHT:} the $Z^\prime$ and $\phi$ mediation introduced in Ref.~\cite{Cheng:2024hvq}. 
}  \label{fig:mixing_diagrams}
\end{figure}

The above searches at general-purpose detectors typically rely on the existence of hard objects, which commonly appear in high-scale mediator models. On the other hand, the dark showers may also be initiated by the neutral particles at the electroweak (EW) scale, such as the $Z$ and Higgs boson $h$ in the SM. They can singly couple to the dark quarks $\psi$ through the operators
\begin{equation}
\label{eq:Zh_operators}
 \large(i H^\dagger  \overset{\leftrightarrow}{D}_\mu H \large)\large(\overline{\psi} \gamma^\mu  \psi\large)+\text{h.c.} ,\quad \text{and} \quad \large(H^\dagger H\large) \large( \overline{\psi}\psi \large),
\end{equation}
after substituting in the Higgs vacuum expectation value (VEV).
These operators can arise from integrating out some mediators that connect the dark sector and the SM sector, for example, heavy EW-doublet dark quarks which couple to the light dark quark and the Higgs field, or a dark $Z'$ or a dark scalar $\phi$ which mainly couples to the dark sector, but mixes with the SM $Z$ or Higgs boson. These possibilities are depicted in Fig.~\ref{fig:mixing_diagrams}. 
The dark $Z'$ and the dark scalar $\phi$ may also have masses around or below the EW scale if their couplings to SM are suppressed, as in the case that they are only induced by mixings with $Z$ and $h$. In that case, the $Z'$ and $\phi$ decays also give important contributions to the dark showers. Their productions are suppressed by the small mixings with $Z$ and $h$, but the branching ratio (BR) of decaying to dark particles can be close to unity. 

The final state particles of the dark showers initiated by the EW-scale portals are expected to be relatively soft. The decay signals can be exploited to suppress the backgrounds if the light dark hadrons are long-lived. However, the experimental searches that rely on hard-object triggers are ineffective in exploring this scenario. A recast of the CMS emerging jet search~\cite{CMS:2018bvr} to the Higgs portal case~\cite{Carrasco:2023loy} showed that the acceptance rate is in the $10^{-4}$ range. Even though meaningful bounds may still be obtained for certain decay portals with $\mathcal{O}$(cm) decay lengths, the search is not optimized for this scenario. The $Z$-portal decay was considered in Ref.~\cite{Cheng:2019yai}, but it was shown that the emerging jet search at the LHC is weaker than the LHCb reach, which is very sensitive to detect displaced decay vertices up to a couple of centimeters. On the other hand, the data scouting and data parking techniques of the CMS experiment~\cite{CMS:2021sch,CMS:2024zhe} are able to trigger on very low-$p_T$ displaced muons, and hence will be powerful probes of the EW-portal dark shower models. For longer decay lengths ($\gtrsim$ meters), the proposed auxiliary detectors such as MATHUSLA~\cite{Chou:2016lxi,MATHUSLA:2022sze}, Codex-b~\cite{Aielli:2019ivi,Aielli:2022awh}, and ANUBIS~\cite{Bauer:2019vqk} will be advantageous. They are expected to capture LLPs decaying inside the detectors from various production mechanisms.

The kinematics of the dark showers initiated by $Z$, $h$ (and $Z'$, $\phi$) depends only on the parameters of the dark sector but not the mediators in contrast to high-scale mediator models. The event rates can be effectively parametrized by the exotic branching ratios of $Z$ and $h$, or the mixing parameters of $Z'$ and $\phi$. Therefore, dark shower searches in these cases can be performed in a simplified approach with only a few additional parameters like the dark hadron masses, their decay lifetimes and modes. Such results allow easy reinterpretations of constraints or reaches of specific UV models. A major motivation of this work is to provide rough evaluations of dark shower searches from EW-scale $Z, Z', h, \phi$ decays at various detectors, including current constraints and future projections. Note that the limits obtained this way will always succumb to large systematics from unknown dark shower dynamics (see, $e.g.$, Ref.~\cite{Cohen:2020afv,Albouy:2022cin}). The fact doesn't contradict our goal of providing quick projections.

If the dark hadrons are light enough to appear in the flavor-changing neutral current (FCNC) decays of SM hadrons, the FCNC decays can also provide probes of the dark sector~\cite{Kamenik:2011vy}. Even if the portal interactions conserve SM flavor, the FCNC amplitudes still inevitably arise at one loop. As the widths of SM FCNC decays are small, the FCNC processes are sensitive to other tiny exotic decays beyond the SM. Furthermore, heavy-flavored QCD hadrons have masses significantly below the EW scale, which implies that not only high-energy experiments but also high-intensity experiments can produce a substantial number of heavy-flavored hadrons~\cite{Kou:2018nap}. The exotic decays of the heavy-flavored hadrons are studied in various experiments~\cite{CMS:2016plw,LHCb:2020frr,Belle-II:2023esi,CHARM:1985anb}. This also allows us to perform simplified projections of dark sector reach. The relevant parameters will be the FCNC branching ratios to the dark sector, and the lifetimes and decay modes of the dark hadrons. These projections have different dependences on the dark sector parameters relative to the dark showers from $Z, Z', h, \phi$ decays, so they provide complementary tests. The comparisons of the experimental reaches from  $Z, Z', h$, and $\phi$ decays and FCNC processes can be made within specific benchmark models. 

The rest of the paper is organized as follows. In Sec.~\ref{sec:Pheno}, we briefly introduce the phenomenology of dark hadrons with EW-portal couplings, including their composition, symmetry, and the effective field theory (EFT) of decays. Sec.~\ref{sec:DarkShower} introduces various collider constraints from dark shower production with LLP signals. The results are shown, including both recasts of current data and projections for future experiments. Similarly, analyses and constraints for dark hadron production in FCNC processes are presented in Sec.~\ref{sec:FCNC}. The above results are combined according to a couple benchmark models in Sec.~\ref{sec:model}, where different production mechanisms are related by UV parameters. We briefly summarize in Sec.~\ref{sec:Summary}. App.~\ref{app:Trigger} describes the dimuon vertex trigger efficiency that we extrapolate for HL-LHC projections. App.~\ref{app:benchmark} contains the details of the two benchmark models used to compare the reaches of various experiments.

\section{Phenomenology of Dark Hadrons}
\label{sec:Pheno}

The low energy degrees of freedom of the dark QCD are dark hadrons. With more than one light flavors, the lightest dark hadrons are expected to be dark pions $\hat{\pi}_a$, the pseudo-Nambu-Goldstone bosons (pNGBs) of chiral symmetry breaking. They likely form the main compositions of the dark shower. They are related to the dark quarks, in the example of two flavors,  by
\begin{equation}
\hat{\pi}_a \sim i (\overline{\psi}_L^{\,\prime} \sigma_a \psi_R^\prime -  \overline{\psi}_R^{\,\prime} \sigma_a \psi_L^\prime) = \overline{\psi}^{\,\prime} \hspace{-0.5mm} i \sigma_a \gamma_5 \psi^\prime \,,
\label{eq:piondefinition}
\end{equation}
where 
$\psi'$ denotes the dark quark mass eigenstates, and $\sigma_a$ are the Pauli matrices. The $\hat{\pi}_{1,3}\; (\hat{\pi}_2)$ have $J^{PC} = 0^{-+}\; ( 0^{--})$~\cite{Cheng:2021kjg}. 

\begin{figure}

\centering
\includegraphics[width=7.2 cm]{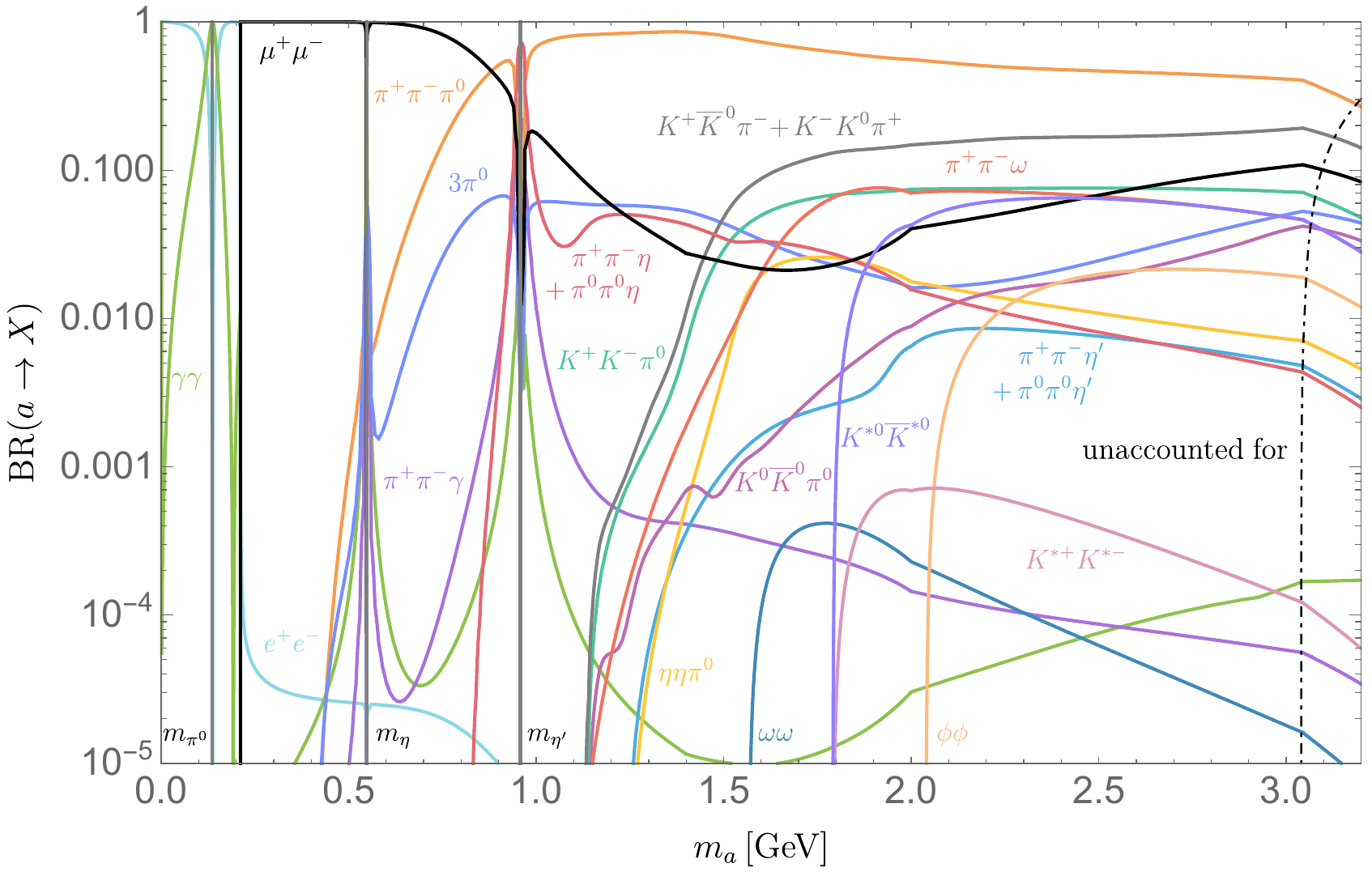}
\includegraphics[width=7.2 cm]{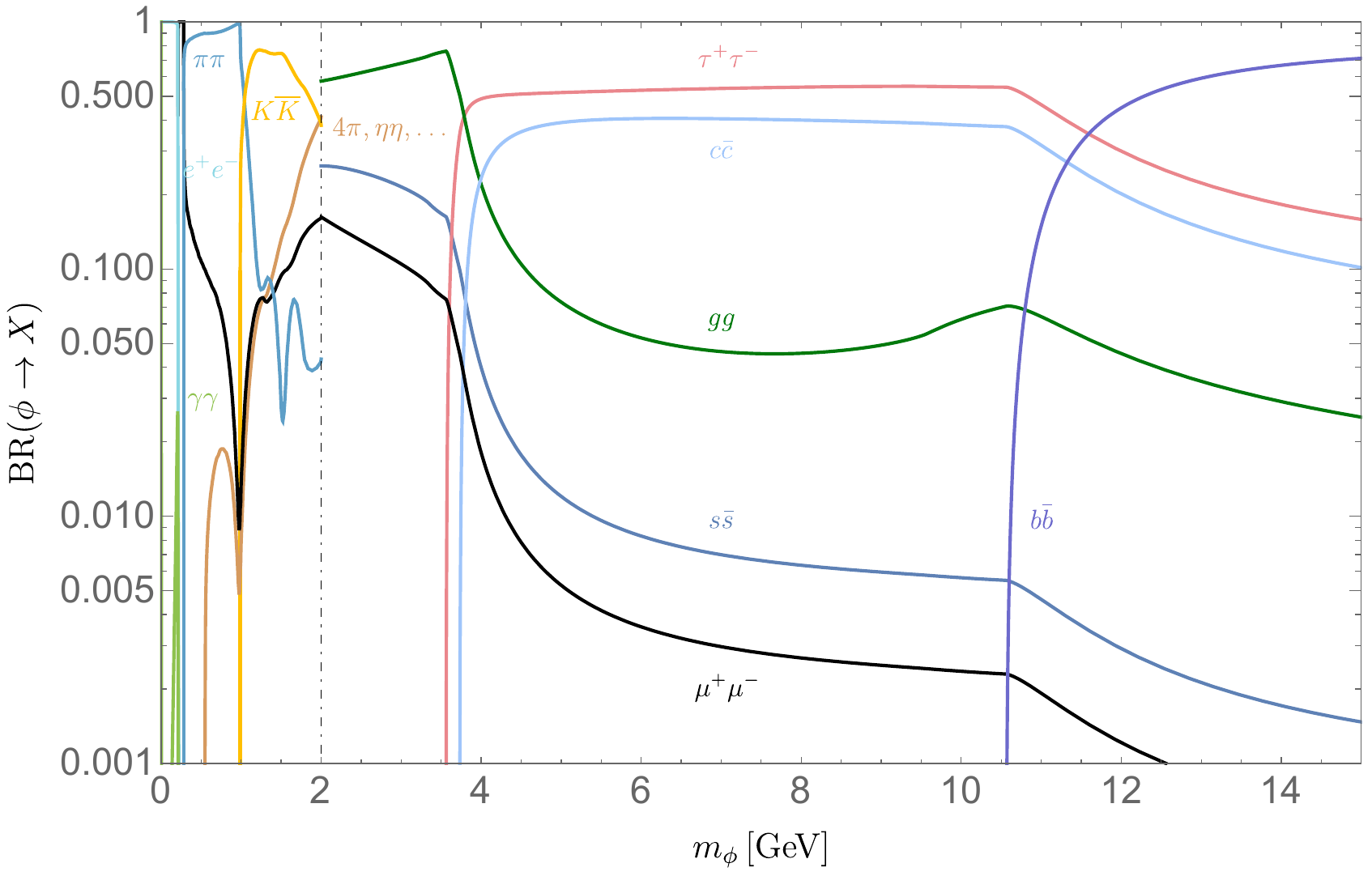}
\includegraphics[width=7.2 cm]{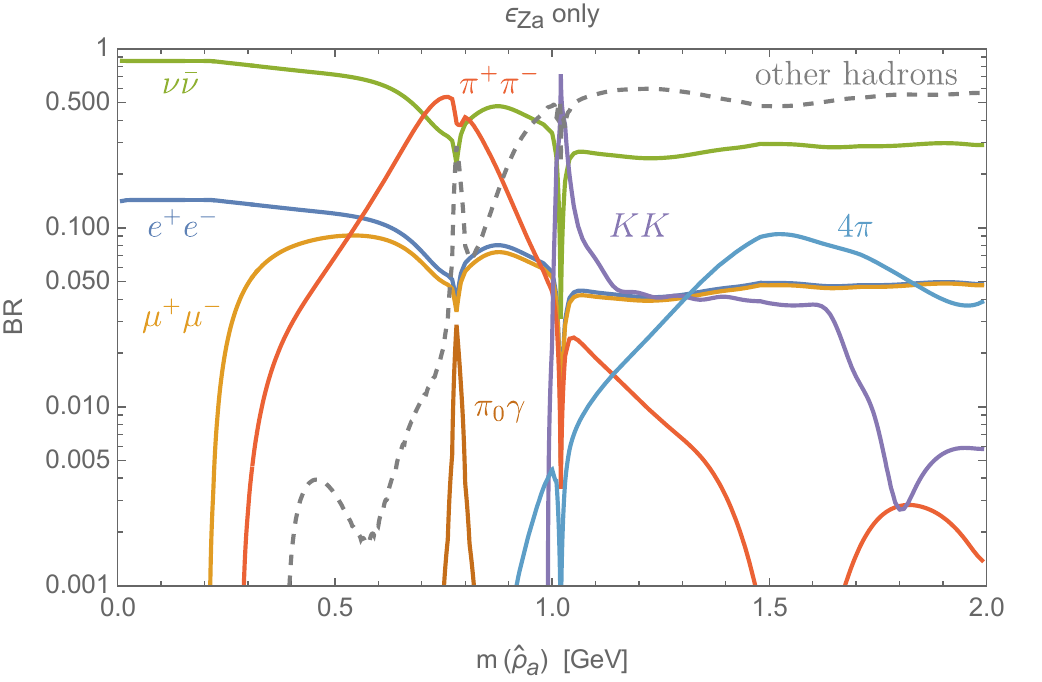}
\caption{{\bf TOP-LEFT:} The branching ratios of a $CP$-odd dark pion decay through the mixing with $Z$ only. { \bf TOP-RIGHT:} The branching ratios of a $CP$-even dark pion decay through the mixing with Higgs only. Both plots are taken from~\cite{Cheng:2021kjg}. {\bf BOTTOM: } The branching ratios of a vector dark meson that mixes with the SM $Z$ boson instead of a photon. The results are obtained from DarkCast~\cite{Baruch:2022esd}.
\label{fig:hadrondecays}}
\end{figure}

Being pseudoscalars with highly suppressed couplings to the SM, the $CP$-odd $\hat{\pi}_{1,3}$ (and $\hat{\pi}_2$ if $CP$ is violated in the dark sector) behave effectively as composite axion-like-particles (ALPs). The effective coupling between a dark pion and SM fermions $f$ can be described by the effective ALP decay constant $f_a^{(b)}$:
\begin{equation}\label{eq:piff}
\mathcal{L}_{\pid f\bar{f}}  = -   \frac{\partial_\mu \pid_b}{f_a^{(b)}} \sum_f a_f \bar{f} \gamma^\mu \gamma_5 f\,,
\end{equation}
where $a_f$ is the axial charge of $f$. This effective coupling is generated by integrating out $Z$ (and $Z'$). 
The effective decay constant $f_a^{(b)}$ can be calculated in terms of UV parameters of a given model, and the dark pion decay constant $f_{\hat{\pi}}$ (which is defined analogously to $f_{\pi}$ in SM).
Assuming that there are no other lighter states in the dark sector, such as a dark photon, the widths and branching ratios of a light dark pion ($\lesssim 3$~GeV) decays to various SM hadron channels were calculated using a data-driven approach~\cite{Cheng:2021kjg} and the result is shown in Fig.~\ref{fig:hadrondecays}.
 In particular, the inverse decay width of a dark pion to a muon pair is given in $m_{\hat{\pi}}$ and $f_a$ as
\begin{equation}
\label{eq:decaywidth}
\Gamma^{-1}_{\hat{\pi}\to \mu^+\mu^-} \simeq 45~\text{cm}~ \bigg(\frac{f_a}{1~\text{PeV}}\bigg)^2\bigg(\frac{1~\text{GeV}}{m_{\hat{\pi}}}\bigg)  \sqrt{\frac{m_{\hat{\pi}}^2}{m_{\hat{\pi}}^2- 4m_\mu^2}}~.
\end{equation}
 The proper lifetimes of dark pions are easily obtained from their widths to dimuon and the corresponding BR in Fig.~\ref{fig:hadrondecays},
 \begin{equation}
 \label{eq:decaylength}
\Gamma^{-1}_ {\hat{\pi}} =  \Gamma^{-1}_{\hat{\pi}\to \mu^+\mu^-} \cdot {\rm Br} (\hat{\pi}\to \mu^+\mu^-).
\end{equation}
 On the other hand, if CP is a good symmetry of the dark sector, a $CP$-even dark pion will decay via the mixing with the Higgs. The branching ratios of its decays to SM final states are also  shown in Fig.~\ref{fig:hadrondecays}. The decay length is typically too long for current LHC detectors, unless the dark pion is heavy enough ($> 3$~GeV) so that decays to SM heavy fermions are open.

Above the dark pions, there are vector mesons ($\hat{\rho}$) in the dark hadron spectrum. They can be interesting if they are not much heavier than the dark pions and their decays to dark pions are forbidden. In that case, they may also decay back to SM states, constituting part of the dark shower. The decay of dark vector mesons in the EW-portal case happens through their mixing with the SM photon or $Z$ boson, with the latter case a characteristic phenomenon of the EW portal. The branching ratios to SM states for a dark vector meson decay through the $Z$ mixing are shown in Fig.~\ref{fig:hadrondecays}, based on the method of Ref.~\cite{Baruch:2022esd}. Since when $m_{\hat{\rho}} > 2 m_{\hat{\pi}}$, all $\hat{\rho}$ decay to $\hat{\pi}$ and leave no significant impact, for these limits we focus on the dark showers made of dark pions. The results can be re-interpreted for the dark vector mesons if one has the knowledge of the multiplicity of the dark vector mesons in the dark shower.


At one loop, the EW-portal interactions of Eq.~\eqref{eq:Zh_operators} also introduce the accompanying ``flavor portal" interactions that produce dark mesons from SM FCNC transitions, such as $B\to K^{(\ast)} \hat{\pi}$, $B_s \to \hat{\pi}\hat{\pi}$, etc. The down-type FCNC transitions receive an enhanced contribution from loops involving a top quark, so they are more relevant. When the momentum transfer is far below all relevant UV scales, the EW portals introduce the following four-fermion interaction at one loop:
\begin{equation}
\mathcal{L}_{\rm eff}^{\rm{FCNC}} \propto  \frac{g^2 }{128 \pi^2} J_{D}^\mu \bar{d}_j \gamma_\mu P_L d_i  \sum_{q\, \in\, u, c, t} V_{q j}^\ast V_{q i} \mathcal{K}_q + \mathrm{h.c.} ~,
\label{eq:FCNC4fermion}
\end{equation}
where $\mathcal{K}_q$ is a model-dependent dimensionless parameter coming from loop diagrams. 
The contribution is dominated by  $q = t$ with $\mathcal{K}_t \gtrsim \mathcal{O}(1)$. The FCNC productions are not directly proportional to the branching fractions of the $Z$ and $h$ decays into the dark sector, and therefore are sensitive to different combinations of model parameters. In this way, they provide complementary probes of the dark sector at lower energies, which are relevant in precision frontier experiments.

For $2m_\mu <m_{\hat{\pi}} \lesssim m_{\eta^\prime}$, $\hat{\pi}$ dominantly decays to two muons, resulting in the striking displaced vertex (DV) signal at colliders. Their narrow resonance peaks and high multiplicity could further suppress the backgrounds~\cite{Pierce:2017taw}. For higher $m_{\hat{\pi}}$, their decay BR to muons drops below $10\%$. In this case, the strategy of looking for dimuon DVs suffers from the lower BR$(\hat{\pi}\to \mu^+\mu^-)$, though it is still useful. On the other hand, constraining dark hadrons with their hadronic decays is more challenging in general, especially at hadron colliders, where QCD backgrounds are dominant. Some hadronic final states like $K^+K^-$ or $K^{\ast 0}\bar{K}^{\ast 0}$ can be fully reconstructed by track information, while other common modes such as $\pi^+\pi^-\pi^0$ involving neutral components are more difficult to reconstruct. We will focus on a few decay modes, especially $\hat{\pi}\to \mu^+\mu^-$ due to its nice phenomenological features.

Recently, to enhance the LHC's discovery potential, a series of auxiliary detectors have been proposed or developed targeting LLPs with $c\tau \gtrsim \mathcal{O}$(m). These detectors include FASER~\cite{Ariga:2018uku}, FASER-2 from the Forward Physics Facility (FPF)~\cite{Feng:2022inv}, Codex-b~\cite{Aielli:2019ivi},
ANUBIS~\cite{Bauer:2019vqk},
and MATHUSLA~\cite{MATHUSLA:2018bqv}. Many auxiliary detectors enjoy strong shielding that removes most of SM backgrounds, so they are sensitive to small signal rates. In addition, an LLP decay inside the effective volume of the auxiliary detectors will be recognized as a signal event even if the final states are not fully reconstructed, making their sensitivities less dependent on  dark hadron decay modes.

\section{Limits on Dark Shower Signals}
\label{sec:DarkShower}

\begin{figure}[t]
\centering
\includegraphics[width=7cm]{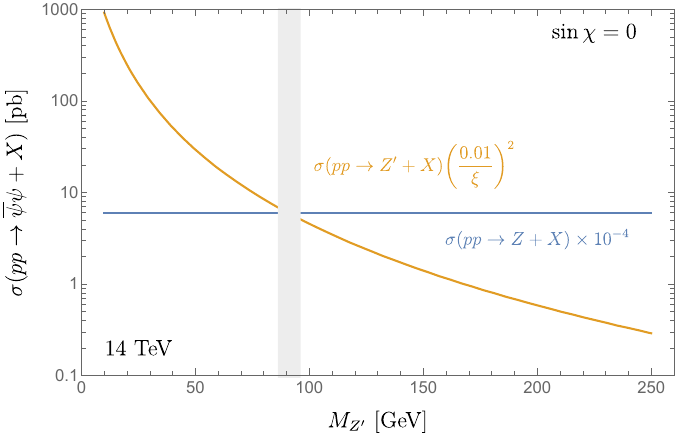}
\includegraphics[width=7cm]{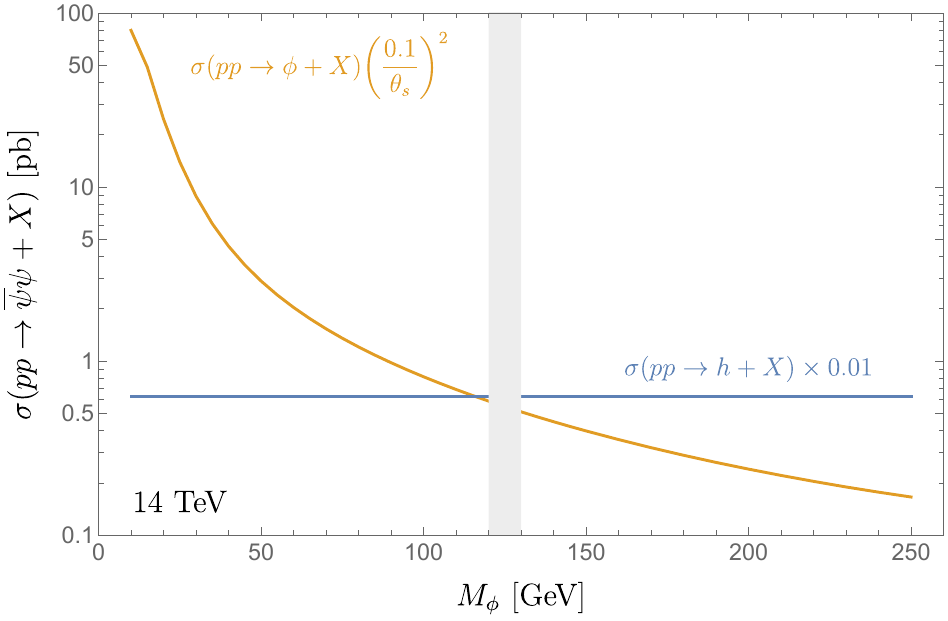}
\caption{Cross section of $Z^\prime$ and $\phi$ production at 14~TeV $pp$ collisions as a function of their masses. \textbf{LEFT:} $Z^\prime$ production with the mixing between $Z$ and $Z'$ taken to be $\xi=0.01$. The blue horizontal line represents the cross section of dark quarks from $Z$ production assuming $\text{Br}(Z\to \bar{\psi}\psi)= 10^{-4}$. \textbf{RIGHT:} $\phi$ production with the $h$-$\phi$ mixing angle $\theta_s=0.1$. For comparison, the blue horizontal line is the dark shower rate from exotic Higgs decays with a BR of 0.01.} 
\label{fig:ppZ2xsec}
\end{figure}

Throughout the paper, we assume a simple dark pion benchmark with two light dark flavors and thus three dark pions. An $SU(3)$ confining dark gauge interaction is assumed, with a gauge coupling of 0.1 around the $Z$ mass. With the EW portal, we consider dark showers at a high energy collider generated from the decays of particles at the EW scale, including SM $Z$, $h$, and also possible dark $Z'$ and scalar $\phi$ which mostly couple to the dark sector.
In this work, we set the simulation for (HL-)LHC with $\sqrt{s}=13(14)$~TeV, respectively. The total production rates of $Z$ and $h$ from $pp$ collisions are~\cite{LHCHiggsCrossSectionWorkingGroup:2016ypw,Ogul:2017zjd,ATLAS:2024nrd}
\begin{equation}\label{eq:Z_h_xsecs} \sigma(pp\to Z) \approx 57.7(62.4)~\mathrm{nb}, \qquad\quad \sigma(pp\to h) \approx 55.7(62.7)~\mathrm{pb}~. 
\end{equation} 
A fraction of $Z$ and $h$ produced may undergo exotic decay to dark quarks $\psi\bar{\psi}$, with their branching ratios constrained by experimental measurements. 
On the other hand, dark $Z^\prime$ and $\phi$ are assumed to couple to SM particles only through mixings with $Z$ and $h$, and hence their productions are suppressed by the mixing angles, but they decay to $\psi\bar{\psi}$ almost exclusively once produced. The cross section of $pp\to Z^\prime \to \psi\bar{\psi}$ for a mixing parameter $\xi=0.01$ (the component of $Z$ in the dark $Z'$ mass eigenstate) as a function of $Z'$ mass is shown in the left panel Fig.~\ref{fig:ppZ2xsec}, following the calculations in Ref.~\cite{Cheng:2024hvq}. The production cross section of the singlet scalar $\phi$ at the LHC, assumed to be produced only from the $\phi-$Higgs mixing, is shown in the right panel of Fig.~\ref{fig:ppZ2xsec}. We adopt the corresponding light beyond the SM (BSM) Higgs value, including all channels~\cite{LHCHiggsCrossSectionWorkingGroup:2016ypw}, multiplied by the benchmark mixing angle squared $\theta_s^2=0.01$. Since the corresponding SM Higgs exotic decay may be sensitive to further model details, we only show the dark shower rate from $h\to \bar \psi{\psi}$ decay with an exotic BR of 0.01 for comparison.  

\begin{figure}
\centering
\includegraphics[width=7cm]{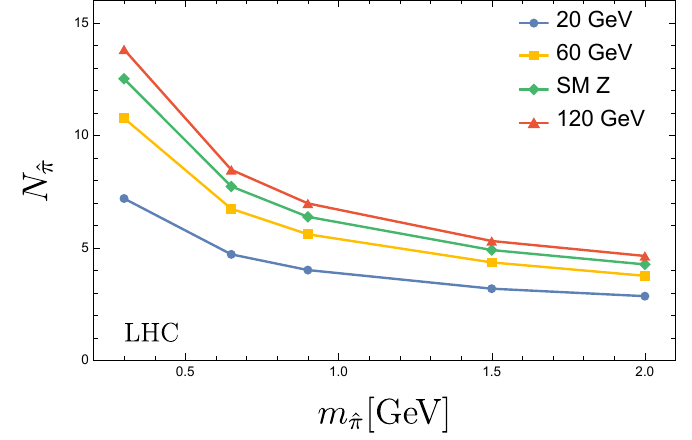}
\includegraphics[width=7cm]{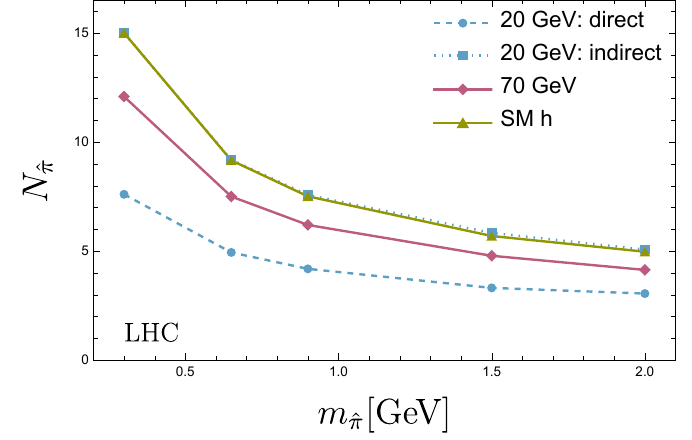}
\caption{Dark pion average multiplicity with different $m_{\hat{\pi}}$ produced by various boson decays to dark shower. \textbf{LEFT:} Distribution for four vector boson benchmarks. \textbf{RIGHT:} Distribution for benchmark scalar boson cases. For the $m_\phi<m_h/2$ benchmark, both the direct production from $pp$ and the indirect production channel from $h\to \psi\bar{\psi}$ decays are plotted.}
\label{fig:statistics}
\end{figure}

\begin{figure}
\centering
\includegraphics[width=5cm]{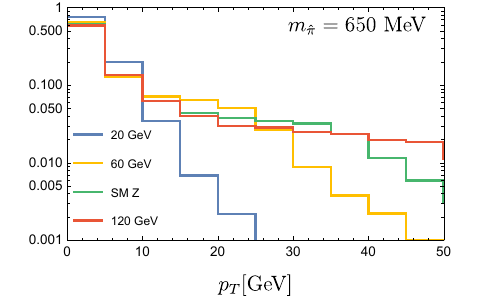}\hspace{-0.4 cm}
\includegraphics[width=5cm]{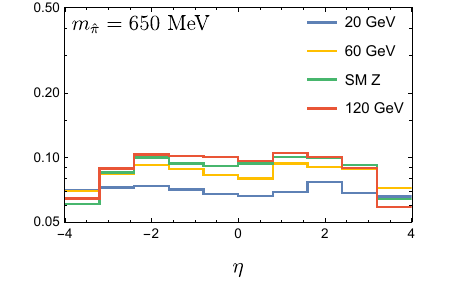}\hspace{-0.4 cm}
\includegraphics[width=5cm]{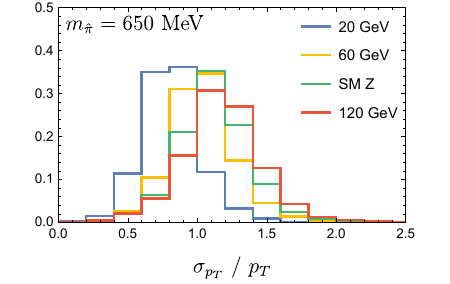}
\caption{The $p_T$, $\eta$ and the normalized standard deviation of $p_T$  distribution of dark pions produced by various vector boson decays to dark shower at the LHC. The $m_{\hat{\pi}}$ is set to the 650~MeV benchmark for all plots.}
\label{fig:kinematics1}
\end{figure}

\begin{figure}
\centering
\includegraphics[width=5cm]{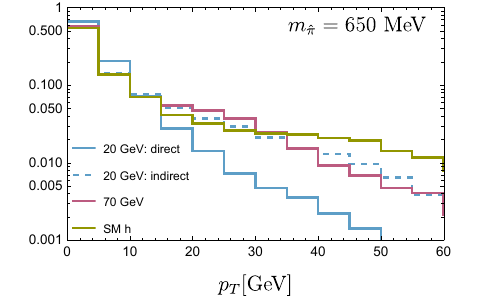}\hspace{-0.4 cm}
\includegraphics[width=5cm]{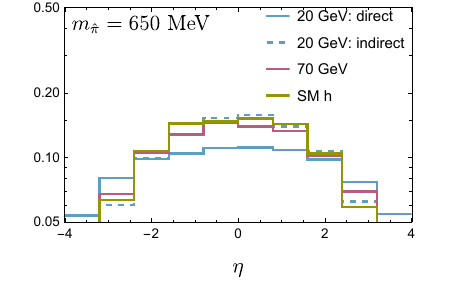}\hspace{-0.4 cm}
\includegraphics[width=5cm]{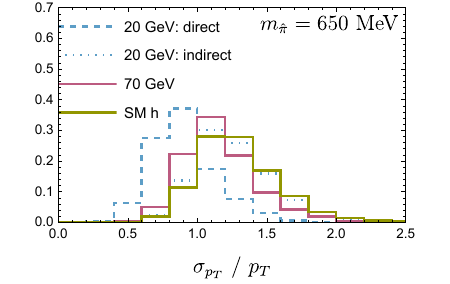}
\caption{Similar to Fig.~\ref{fig:kinematics1} but for scalar bosons.}
\label{fig:kinematics2}
\end{figure}

The Hidden Valley module~\cite{Carloni:2010tw,Carloni:2011kk} of Pythia8~\cite{Sjostrand:2014zea} is employed to simulate various dark shower benchmark samples from heavy boson decays. The dark hadronization is simulated with minimal dark pion multiplicity.\footnote{The choice corresponds to changing the dimensionless parameters aLund to 0.1 and bmqv2 to 2.0 in the Pythia8 Hidden Valley module. The other parameters are kept as their default values. Though the smaller dark pion multiplicity from this benchmark value reduces the overall signal rate, the overall signal efficiency is compensated by the fact that the average dark pion energy becomes higher, leaving the final result less sensitive to these parameters.} The dark pion kinetic distributions such as $p_T$, $\eta$, and multiplicity of dark hadrons are crucial for our analysis~\cite{Albouy:2022cin}. Fig.~\ref{fig:statistics} displays the average dark pion multiplicities in the dark showers from $Z, Z', h, \phi$ decays with different $m_{\hat{\pi}}$. For $m_\phi<m_h/2$, the light $\phi$ can be produced either from $pp$ collision directly or indirectly from $h\to \phi\phi$ decays. For all benchmark EW-portal models, the average dark pion multiplicity $N_{\hat\pi}$ drops with $m_{\hat{\pi}}$ with $N_{\hat\pi}\propto m_{\hat{\pi}}^{-0.6}$ approximately. The simulated dark pion kinematic distributions from various initial vector boson benchmarks are shown in Fig.~\ref{fig:kinematics1}, while all scalar boson cases are plotted in Fig~\ref{fig:kinematics2}. In all cases, the majority of dark pions are soft ($p_T \lesssim 10$~GeV), while the high $p_T$ tail strongly depends on the initial boson mass. Since the initial bosons are also moderately boosted by QCD emission, the maximum dark pion $p_T$ may exceed half of the initial boson mass. The small average dark pion $p_T$ thus reduces the signal efficiency for searches only sensitive to hard tracks. The average $|\eta|$ of dark pions decreases with increasing the initial boson mass. Also, dark pions produced from $Z^{(\prime)}$ have higher $|\eta|$ when compared to those from scalar bosons with a similar mass. This is because the major production channel for $Z^{(\prime)}$ is $q\bar{q}$ with highly asymmetric parton distribution functions (PDFs), in contrast to the dominant $gg$ fusion in the scalar cases. In practice, the $\eta$ distribution in the high $|\eta|$ tail is directly related to the performance of experiments like LHCb. We also plot the spread of dark pion $p_T$ within a single event in Fig.~\ref{fig:kinematics1}. In particular, the plotted dimensionless quantity is the standard deviation of dark pion $p_T$ divided by their average in each event. Generically, as the initial boson mass increases, generating more dark pions in the event, the dark pion $p_{T}$ are more unevenly distributed in the event.
 
For later studies, we take a benchmark $m_{Z^\prime}=20$~GeV for the dark $Z'$ such that the dark pion kinematics and limits will be significantly different from that of $Z$ and not light enough to be tightly constrained by flavor factory searches (e.g., Ref.~\cite{Campajola:2021pdl}). 
For the scalar $\phi$, we see that the indirect production $h\to \phi\phi$ for a 20~GeV $\phi$ has very similar $N_{\hat{\pi}}$ and $p_T$ distributions (for low and moderate $p_T$) as the $h\to \psi\bar{\psi}$. Therefore, we expect similar reaches as the dark showers induced directly from the Higgs decay.\footnote{The $p_T$ distribution differences within each event may affect the 2DV searches.} This does not hold for all $m_\phi$, but the result is not expected to change a lot due to the compensating effects between $N_{\hat{\pi}}$ and $p_T$. To avoid two different production mechanisms whose ratio is highly model-dependent, we take a value $m_\phi=70$~GeV so that $h\to \phi \phi$ decay is not open for our benchmark study. It also represents a mass value intermediate between 20 GeV and $M_Z, M_h$. The branching ratios of both $Z^\prime$ and $\phi$ to dark shower are set to unity. Finally, we assume that all dark pions in the dark shower have identical lifetimes for simplicity.

\subsection{Dark Shower Sensitivity at CMS}
\label{ssec:darkshower_main}
In Ref.~\cite{CMS:2021sch}, the data scouting technique~\cite{Carloni:2011kk} is applied to probe low mass dimuon displaced vertices (DVs) using the CMS data. The data scouting technique reduces the trigger threshold online, significantly benefiting the search for low-mass dimuon resonances. In this context, the data of muon pairs that pass the lower-level triggers are recorded, with the data flow per event greatly reduced to accommodate a higher trigger rate.

Given that the analysis of Ref.~\cite{CMS:2021sch} targets different models, it is essential to recast the publicly available data for dark shower searches. The procedure follows a similar approach in Ref.~\cite{Born:2023vll}. A set of truth-level cuts is first applied on each dimuon pair resulting from dark hadron decays. Only the muon pairs with both $p_{T,\mu}>3$~GeV, $|\eta_\mu|<2.4$ are selected. In addition, the DV's transverse displacement $l_{xy}$ must be $<$11~cm. The angle ($\Delta \phi_{\hat{\pi}}$) between the DV's momentum and its spatial displacement from the primary vertex (PV) must be less than 0.02. When $m_{\hat{\pi}}$ is small, the DV's invariant mass window primarily results from the finite resolution of the tracker, which is about 1.1\% of the resonance mass. 
We therefore take the window to be $m_{\mu\mu}\in[0.978 m_{\hat{\pi}},1.022 m_{\hat{\pi}}] $ to maximize sensitivity.\footnote{Due to final state  photon radiation, the reconstructed $m_{\mu\mu}$ sometimes lie outside the window, further reducing the signal efficiency. The effect is more significant for heavier $m_{\hat{\pi}}$, leading to a maximum signal efficiency loss $\sim 25 \%$, which is included in the final results.} To obtain the final signal yield, each remaining candidate DV will be weighed by the trigger and selection efficiencies reported in Ref.~\cite{CMS:2021sch} according to their properties, such as their $p_T$ or $l_{xy}$. As dark pions are likely to decay back to SM particles within a small solid angle, no isolation requirements are placed on candidate DVs.\footnote{A similar analysis in Ref.~\cite{Born:2023vll} demonstrates that the isolation requirements on muons will not significantly affect the final exclusion limit.} The signal yield is divided into several $l_{xy}$ bins. Typically, the results are dominated by one bin, so only the maximal signal significance from the dominating $l_{xy}$ bin is adopted to get the limit instead of combining different bins.

Since the CMS reported backgrounds change very slowly with $m_{\mu\mu}$ away from SM resonances, we calculate the background yield based on the average background yield per $m_{\mu\mu}$ provided and the $m_{\mu\mu}$ mass window defined above. To better match with the reported maximum signal yields reported, the background yield in each $l_{xy}$ bin is assumed to have a systematic uncertainty. A 3\% relative background fluctuation in each $l_{xy}$ bin is assumed, which leads to results compatible with the signal event upper limit reported by CMS. The 95\% C.L. limits on the exotic BRs are obtained by solving the approximate relation
\begin{equation}
\frac{S_{95}}{\sqrt{S_{95}+B+(\Delta B)^2}}=2~,
\end{equation}
where $S_{95}$, $B$ and $\Delta B$ stand for the signal yield at 95\% C.L., the background yield and the systematic background uncertainty in the dominating $l_{xy}$ bin, respectively. 

\begin{figure}[h!]
\includegraphics[height=5 cm]{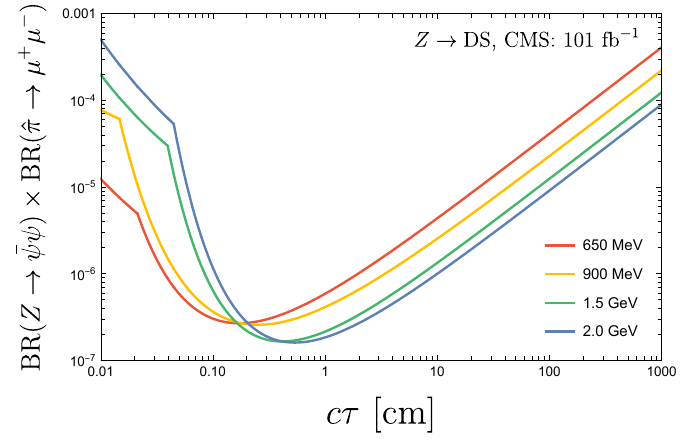}
\includegraphics[height=5 cm]{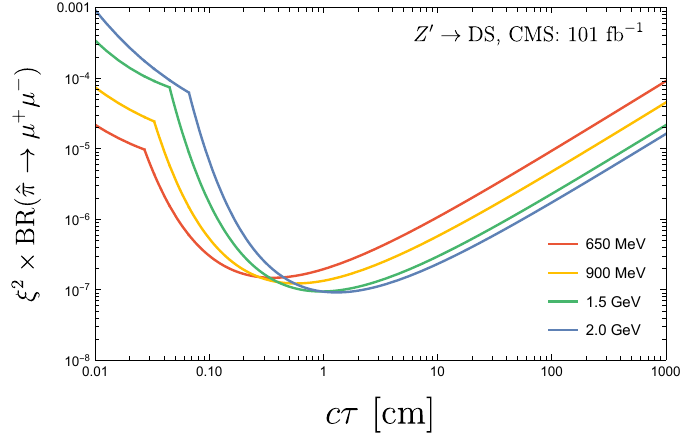}
\includegraphics[height=5 cm]{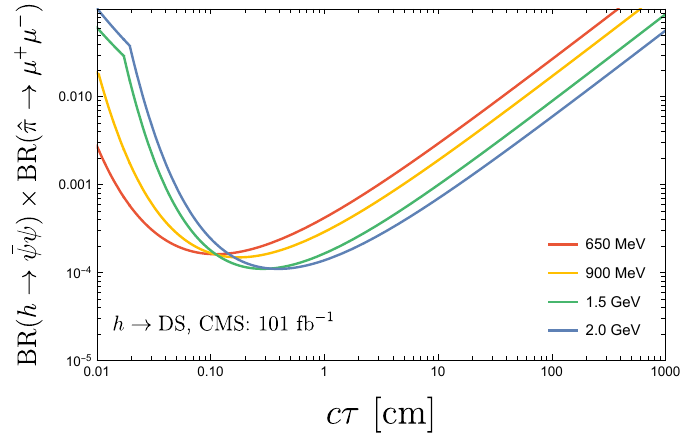}
\includegraphics[height=5 cm]{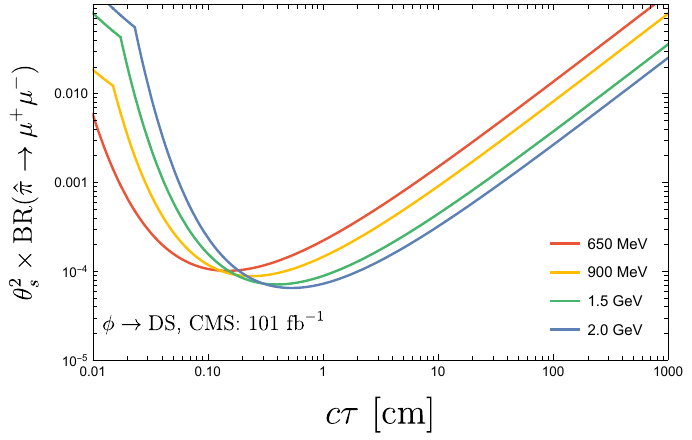}
\caption{The 95\% C.L. limits on $Z$, $Z^{\prime}$, $h$, and $\phi$-initiated dark shower by recasting current CMS dimuon DV search~\cite{CMS:2021sch}. The different colors correspond to several $m_{\hat{\pi}}$ benchmark values. All limits are modified by BR$(\hat{\pi}\to \mu^+\mu^-)$. The kinks in the small $c\tau$ region are induced by the dominant $l_{xy}$ bin switch  between 0.2-1 and 1-2.4~cm, which are of different behaviors. \textbf{TOP-LEFT:} The current CMS limits on the product of the exotic BR$(Z\to \psi\bar{\psi})$. \textbf{TOP-RIGHT:} The current limit of a $Z^\prime$ that mainly decays to dark shower with $m_{Z^\prime}=20$~GeV. The limits are given in terms of $\xi^2 $. \textbf{BOTTOM-LEFT:} The limit on exotic Higgs BR($h\to \psi\bar{\psi}$). \textbf{BOTTOM-RIGHT:} The current limit of the light scalar $\phi$ that mainly decays to dark shower with $m_{\phi}=70$~GeV. The limits are given in terms of its Higgs mixing angle squared $\theta_s^2 $.}
\label{fig:Shower_CMS_1}
\end{figure}

Since the above procedure works for a wide range of initial boson masses, they can be applied to dark showers initiated by various bosons, with results shown in Fig.~\ref{fig:Shower_CMS_1}.
 All limits are delivered as functions of $c\tau(\hat{\pi})$ and $m_{\hat{\pi}}$ only. For the SM initial bosons $Z$ and $h$, the limits are set on their exotic BR to dark shower. Limits from BSM initial bosons $Z^\prime$ and $\phi$ are given as their mixing parameters $\xi^2$ or $\theta_s^2$. The limits from all four cases are scaled by BR($\hat{\pi}\to \mu^+\mu^-$) due to the nature of 1DV searches. Conversely, if more than one dimuon DVs with the same mass are found in a single event, backgrounds from flavored hadrons and detector effects shall diminish as it is unlikely to give two DVs with the same mass this way. However, with the data scouting approach, the muon $p_T>3$~GeV cut leads to low DV efficiency of $\mathcal{O}(10^{-3})$. The chances of reconstructing more than one DV in an event are too low to give useful constraints with the current data, but could be relevant at the HL-LHC discussed later.

All results in Fig.~\ref{fig:Shower_CMS_1} share qualitative features. The optimal $c\tau(\hat{\pi})$ of $\mathcal{O}$(cm) as the dark pion distributions shown in Fig.~\ref{fig:statistics} and~\ref{fig:kinematics1} are similar to each other. Such optimal $c\tau(\hat{\pi})$ becomes larger for a heavier dark pion since the average transverse boost decreases with larger $m_{\hat{\pi}}$. The limit on Higgs exotic BR limits are $\sim 10^3$ weaker than those of $Z$ due to the LHC rate differences in Eq.~\eqref{eq:Z_h_xsecs}. The ratio of $\sim 10^3$ is also found in the mixing parameter constraints between the 20~GeV $Z^\prime$ and 70~GeV $\phi$ coincidentally. We also note that a 20 GeV $Z'$ has a production cross section $\sim 300$ times larger than one with a mass equal to $M_Z$, but the reach is only a few times better due to a lower $N_{\hat{\pi}}$ and a smaller $p_T$ distribution reducing the signal efficiency. This implies that the experimental reach will not be a sharp function of $M_{Z'}$.

For dark pions with longer decay lengths, the dark shower can also be searched at the muon detectors if it produces high multiplicity clusters of hits~\cite{ 
CMS:2024bvl}. The current search requires a large missing $p_T$, $p_T^{\rm miss} >200$~GeV, which reduces the signal acceptance down to $\sim 1\%$. 
The reaches in the branching ratios of the Higgs boson decaying into dark showers are about $10^{-2}$ at the optimal decay lengths around 1 m, depending on the compositions of the final states. However, the muon final states leave too few hits and cannot be used. Therefore, it would not apply to the ($CP$-odd) dark pions with a mass between $2m_\mu$ and 900 MeV where the dimuon is the dominant decay channel.

\subsubsection{Future Projections}

Employing the averaged-background-level method above, it is then possible to project the HL-LHC limit using the dimuon scouting approach. Both signal and background are scaled up according to the integrated luminosity, increasing from 101 to 3000~fb$^{-1}$. Furthermore, due to the implementation of advanced technologies and algorithms in the high-luminosity phase, the trigger efficiency and acceptance of dimuon DVs are expected to improve. Specifically, we expect that the $l_{xy}<11$~cm requirement for DVs can be extended to about 90~cm, where muon tracks will intersect at least four two-strip sensor layers~\cite{Contardo:2015bmq}. The corresponding DV trigger efficiency, which declines rapidly with $l_{xy}$ in Ref.~\cite{CMS:2021sch}, is anticipated to be higher at the HL-LHC. In App.~\ref{app:Trigger}, we postulate several scenarios of trigger efficiency based on current CMS values and track reconstruction for high impact parameter tracks ($e.g.$ Ref.~\cite{ATLAS:2017zsd}). For the detector's reconstruction efficiency after triggering, we take the conservative estimate of 70\% for all dimuon DVs, consistent with the results in Ref.~\cite{CMS:2021sch}. The signal yields beyond $l_{xy}=11$~cm are further divided into three bins with boundaries of 11, 20, 40, and 90~cm.

Besides the signal efficiency, another key element for HL-LHC projection is the background mitigation. Due to the significant uncertainties from the upgraded detector performance and higher pile-up level, providing a precise background prediction across $m_{\mu\mu}$ and $l_{xy}$ is impractical without experimental studies. Instead, the approximate range of backgrounds may be estimated with the released data and some assumptions. The background $l_{xy}$ distribution in Ref.~\cite{CMS:2021sch} shows that the SM backgrounds originate from at least two sources. When $l_{xy}\lesssim 3$~cm, the background drops exponentially with $l_{xy}$, indicating that the dominant source is the decay of flavored hadrons in the SM. For larger $l_{xy}$, the background drop follows an approximate power law of $l_{xy}^{-2.5}$, suggesting that for $l_{xy}$ much larger than the bottom quark lifetime, the backgrounds are dominated by the accidental crossing of real and fake tracks. 

\begin{figure}[h!]
\includegraphics[height=5 cm]{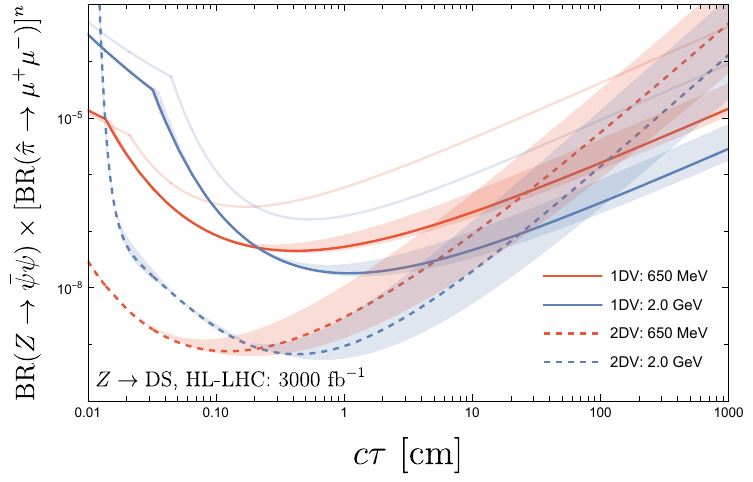}
\includegraphics[height=5 cm]{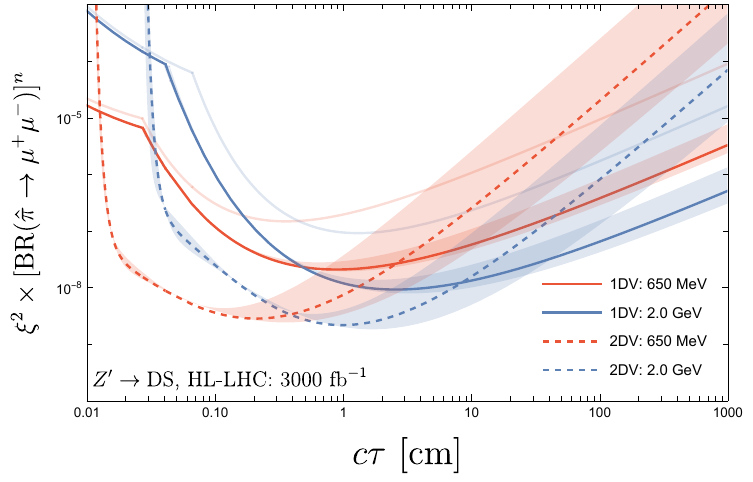}
\includegraphics[height=5 cm]{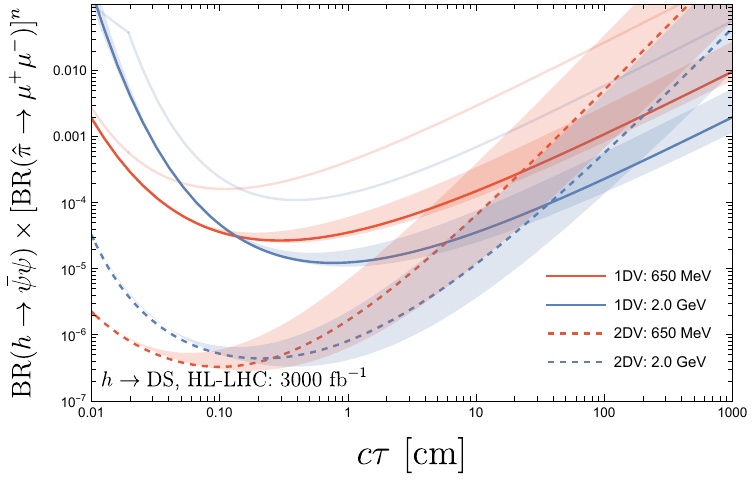}
\includegraphics[height=5 cm]{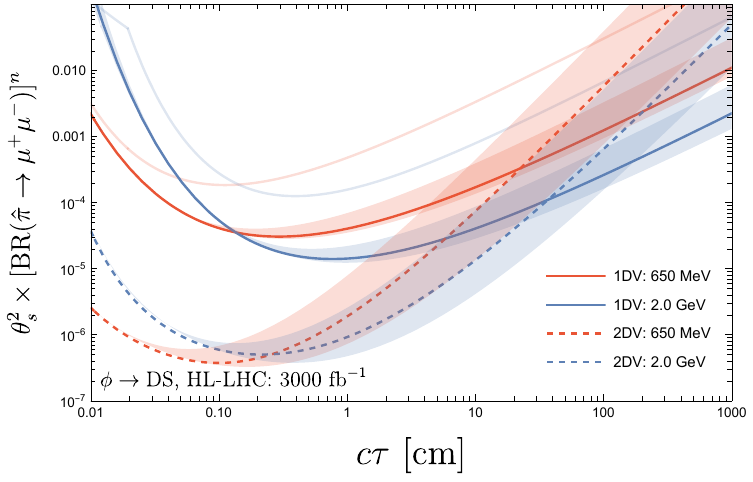}
\caption{Similar to Fig.~\ref{fig:Shower_CMS_1} but the projected dark shower limits from dimuon scouting at the HL-LHC. The different colors correspond to $m_{\hat{\pi}}$ benchmark values, while the light-shaded curves corresponds to current limits in Fig.~\ref{fig:Shower_CMS_1}. The uncertainties due to varying trigger efficiency scenarios are shown in bands, with the solid contours in the center stand for limits in the median trigger efficiency scenario in App.~\ref{app:Trigger}. The 2DV limits are also plotted as dashed curves, corresponding to the median trigger efficiency scenario with uncertainties shown as bands. On the label of the vertical axis, the power of the dimuon BR $n=1(2)$ for 1(2)DV limits.
}
\label{fig:Shower_CMS_2}
\end{figure}

The dimuon DV background at HL-LHC is estimated based on current data. In particular, the background is first rescaled proportional to luminosity. The $l_{xy}<11$~cm part is proportional to the CMS data, while the $l_{xy}\in (11,90]$~cm part is estimated from the yields in $l_{xy}\in [7,11]$~cm bin, assuming the backgrounds drop as $l_{xy}^{-2.5}$. We also assume that the background yield in each $l_{xy}$ bin will also increase by the same factor as the improvement of the average signal efficiency described in App.~\ref{app:Trigger}.\footnote{We also assume the impact from higher pile-ups in high-luminosity runs will be compensated by the timing layer~\cite{Butler:2019rpu} and no extra factors are applied.} A 3\% background systematic uncertainty is applied to all $l_{xy}$ bins. The HL-LHC projections from such background estimations of dimuon scouting are plotted in Fig.~\ref{fig:Shower_CMS_2}. The different trigger efficiency scenarios are translated to the uncertainty band, with the contours in the middle corresponding to the median scenario. The improvement from LHC to HL-LHC era is most significant when the dark pion lifetime is greater than 10~cm due to the signal $l_{xy}$ acceptance range is extended from 11 to 90 cm. In contrast, limits for the short lifetime regime suffer from significant background systematic uncertainty and show little improvement. 

In the scouting search, the macroscopic decay length of dark mesons is the primary feature that separates the signal from SM backgrounds. As the chance of finding two DVs in an event is suppressed by the single DV efficiency twice, the LHC limit from Run 2 is not as competitive. However, searching for more than one DVs may become possible at a higher luminosity. The excellent dimuon mass resolution referred to earlier implies that, even in the HL-LHC era, events featuring two dimuon DVs that fall into the same mass window will be exceedingly rare, thus legitimizing the background-free assumption. To derive the 2DV constraints on HL-LHC from data scouting conservatively, we first assume 
the overall efficiency of the event is the product of two individual DVs' efficiencies.\footnote{Very often, in events having a dark shower, there are more than two dark hadrons that decay to the dimuon final state. Only the two DVs with the highest trigger efficiencies are considered. Numerically, we found this approach to be a good approximation since the DV efficiencies in an event are often in a hierarchical structure, dominated by the ones with proper lifetime and high $p_T$.} In addition, all DVs with $l_{xy}<1$~cm are dropped to avoid introducing flavored backgrounds. Based on the assumption of background and Poisson statistics, the 95\% C.L. limit of a background-free signal region corresponds to a signal yield $\sim$ 3. We plot the projected 95\% C.L. HL-LHC 2DV constraints on all benchmarks in Fig.~\ref{fig:Shower_CMS_2} as dashed curves, which need to be scaled by BR$(\hat{\pi}\to \mu^+\mu^-)^2$. As expected, the 2DV constraints derived this way are most relevant when the dark hadron's lifetime makes their trigger efficiencies optimal. The constraints weaken much faster than the 1DV case for both smaller and larger lifetimes. Notably, the difference between 1DV and 2DV $Z^\prime$ initiated dark shower limits is much smaller than in the other cases, which is probably caused by the low $N_{\hat{\pi}}$ and more evenly distributed $p_T$ among dark pions.

The 2DV projections in Fig.~\ref{fig:Shower_CMS_2} are conservative, limited by the square of the small trigger efficiency for a single soft DV. Note that the data scouting procedure only needs to trigger on a pair of muons, which do not have to share the same vertex. The actual trigger rate of an event with two or more dimuon DVs will be higher than the above estimation, though the general reconstruction efficiency of muons is unavailable. Moreover, improvements in the hardware and trigger algorithms are expected at the HL-LHC, which render the multi-DV searches more viable. Despite all the arguments suggesting much stronger 2DV limits, we only report the most conservative 2DV constraints in this work until a more realistic projection of the signal efficiency and background assessment at the HL-LHC becomes available.

\subsection{Dark Shower Sensitivity at LHCb}
\label{ssec:shower_LHCb}
\begin{figure}[h!]
\centering
\includegraphics[width=7 cm]{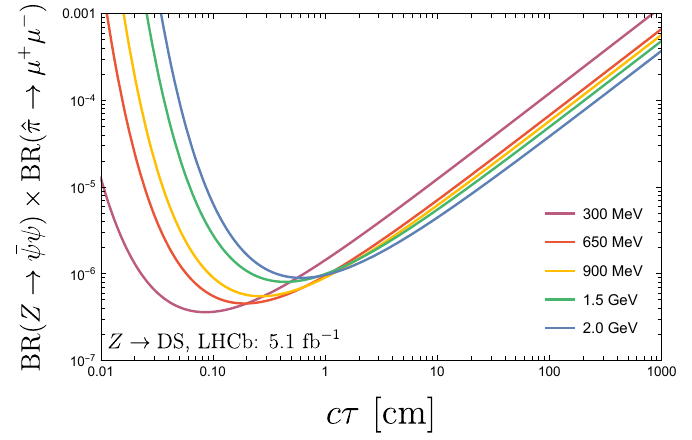}
\includegraphics[width=7 cm]{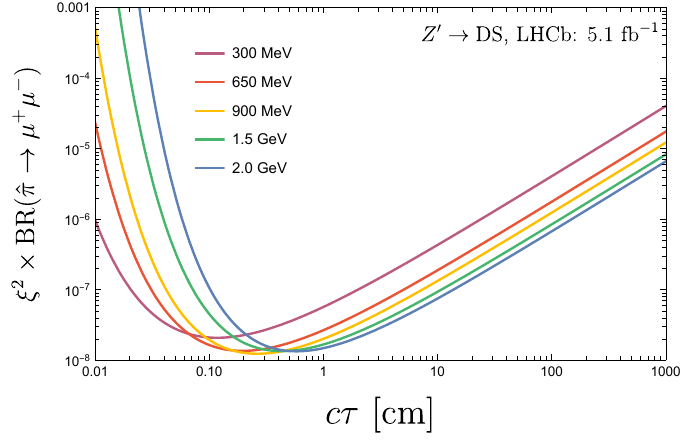}
\includegraphics[width=7 cm]{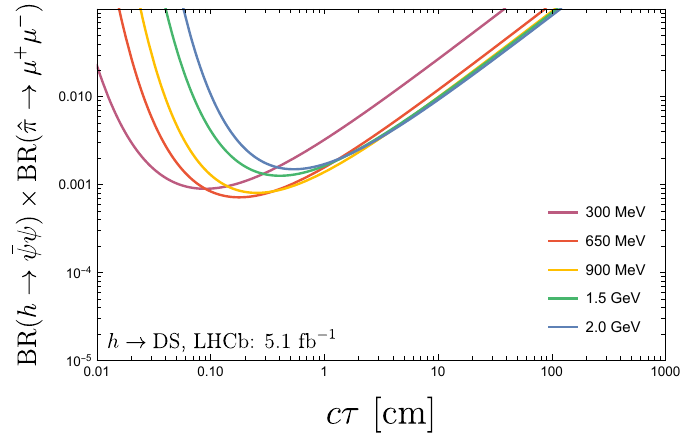}
\includegraphics[width=7 cm]{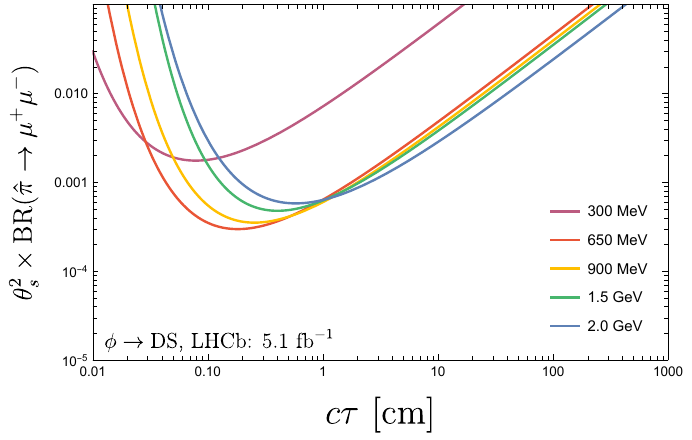}
\caption{The 95\% C.L. dark shower limits from recasting current LHCb dimuon DV search~\cite{LHCb:2020ysn}. The format is similar to Fig.~\ref{fig:Shower_CMS_1}. 
}
\label{fig:Shower_LHCb_1}
\end{figure}

The LHCb detector with low trigger thresholds and higher vertex resolutions is another powerful tool to look for dimuon DVs from dark showers, especially for $c\tau(\hat{\pi})\in[0.1,1]$~cm case, which matches the LHCb vertex locator (VELO) size. To recast recent LHCb dimuon DV search results, one can follow the analysis in Ref.~\cite{LHCb:2020ysn}. In particular, we apply at truth level the displaced search cuts listed in Table~1 of Ref.~\cite{LHCb:2020ysn} and compare to the cross section limits for promptly-produced $X\to \mu^+ \mu^-$~\cite{LHCb:2020ysn}. The events yield in 8 $m_{\mu\mu}$ bins around $m_{\hat{\pi}}$, each representing half of $m_{\mu\mu}$ resolution at LHCb, are summed to obtain the background.\footnote{Similar to the CMS case, the background events in each bin are averaged between bins with $m_{\mu\mu}\in[m_{\hat{\pi}}-100\text{MeV},m_{\hat{\pi}}+100\text{MeV}]$ to stabilize the prediction.} No background systematic uncertainties are assumed since the overall background level is very small already. All current limits for the four benchmark cases are shown in Fig.~\ref{fig:Shower_LHCb_1}, which share the similar behavior of the CMS limits with comparable limits but smaller optimal lifetimes. While being slightly weaker in $Z$, $h$, and $\phi$ cases, the limits for a light $Z^\prime$ are stronger due to the combination of softer muon kinematic cuts at LHCb and the high dark pion $|\eta|$ from $Z^\prime$ decays.

\subsubsection{Future Projections}

\begin{figure}[h!]
\centering
\includegraphics[width=7 cm]{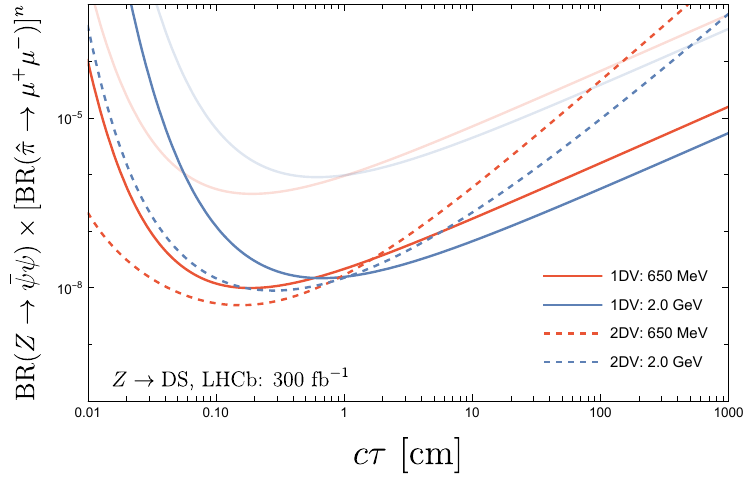}
\includegraphics[width=7 cm]{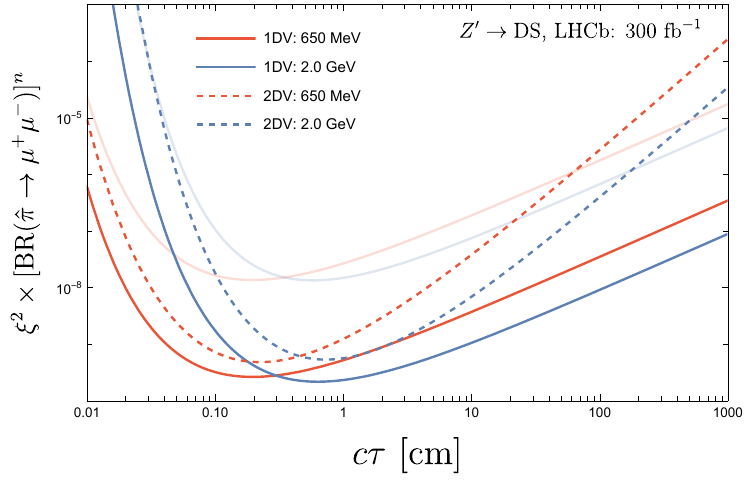}
\includegraphics[width=7 cm]{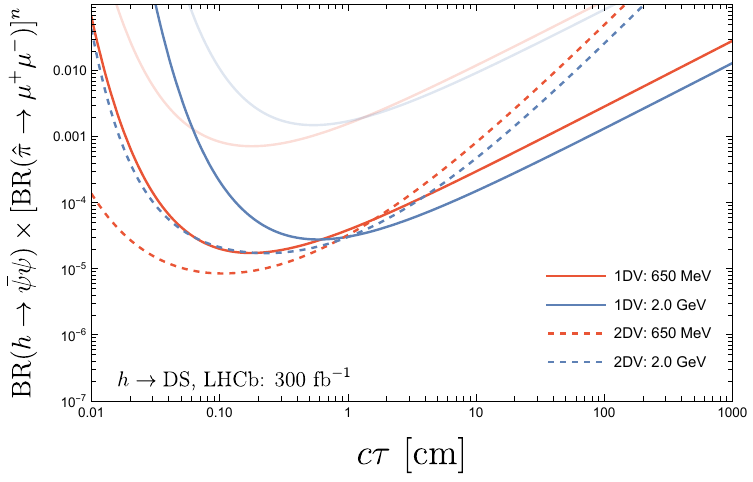}
\includegraphics[width=7 cm]{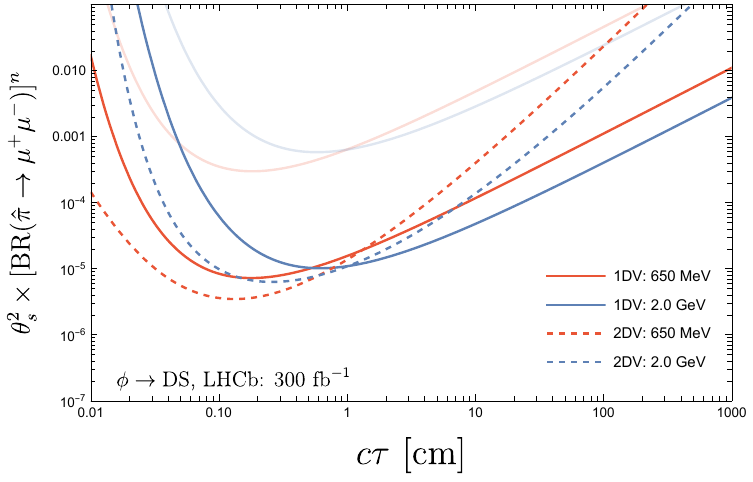}
\caption{Similar to Fig.~\ref{fig:Shower_LHCb_1} but projected for the HL-LHC era. The light-shaded curves stand for current constraints.}
\label{fig:Shower_LHCb_2}
\end{figure}

Owing to an even lower trigger threshold starting from Run 3~\cite{LHCbCollaboration:2014vzo} and the comparatively low pile-up level achieved, we expect the projected dark shower constraints from dimuon DVs for LHCb in run 3 or the high luminosity phase to become stronger than a mere luminosity rescaling. To project the dark shower limit in the high luminosity era, we follow the approach described in Ref.~\cite{Ilten:2016tkc}.  
In particular, the cut on the dimuon DV is softened so that both muon tracks only need to have $p_T>0.5$~GeV, $|p|>10$~GeV, and $\eta\in[2,5]$. The DVs also need to satisfy its total $p_T>1$~GeV and $l_{xy}\in[6,22]$~mm. The future LHCb background in a $m_{\mu\mu}$ window is then rescaled from that of Ref.~\cite{LHCb:2020ysn} with the same magnification factor as the signal efficiency improvement. In addition, due to a potential $m_{\mu\mu}$ resolution improvement, the $m_{\mu\mu}$ mass window width is taken to be 16~MeV ($0.016\times m_{\mu\mu}$) when $m_{\mu\mu} \leqslant(>)$1~GeV~\cite{Ilten:2016tkc}, which further decreases the background. The background deduced this way is compatible with Ref.~\cite{Ilten:2016tkc}. Finally for the detector efficiency, in a previous study (Ref.~\cite{Cheng:2021kjg}) it was found that a detector efficiency in the range $[0.4, 0.8]$ can reproduce the LHCb results~\cite{LHCb:2020ysn}. For simplicity, we apply a global detector efficiency factor of 0.7 for all benchmark signals, with the optimistic hope that it will not be worse at the future LHCb runs. It is also comparable to though slightly higher than the one used in Ref.~\cite{Ilten:2016tkc}. The resulting 1DV projections are shown in Fig.~\ref{fig:Shower_LHCb_2}.

The 2DV result is also shown in Fig.~\ref{fig:Shower_LHCb_2}, following the similar approach described in Sec.~\ref{ssec:darkshower_main}. In particular, an event must have two or more dimuon DVs, each satisfying the kinematic cuts. The global detector efficiency of 0.7 on each DV applies multiplicatively. We also adopt the background-free approximation, given that the background level at LHCb will be much smaller than that of CMS. In contrast to the CMS 2DV limits, the 2DV reaches at LHCb are largely free from the small trigger efficiency but mainly limited by the kinematic requirements, therefore being less suppressed. 
However, due to the low background level at LHCb, the 2DV limits can only compete with the 1DV results for very short dark pion decay lengths ($\lesssim$ few mm).
. 

Besides the dileptonic final states, LHCb also holds the potential for probing displaced hadronic decays of long-lived dark hadrons with signal efficiency approaching that of dimuon cases~\cite{Craik:2022riw}. On the other hand, despite the advanced tracking system and reduced pile-up, significant combinatorial backgrounds may still be present at LHCb.

\begin{figure}
\centering
\includegraphics[width=7 cm]{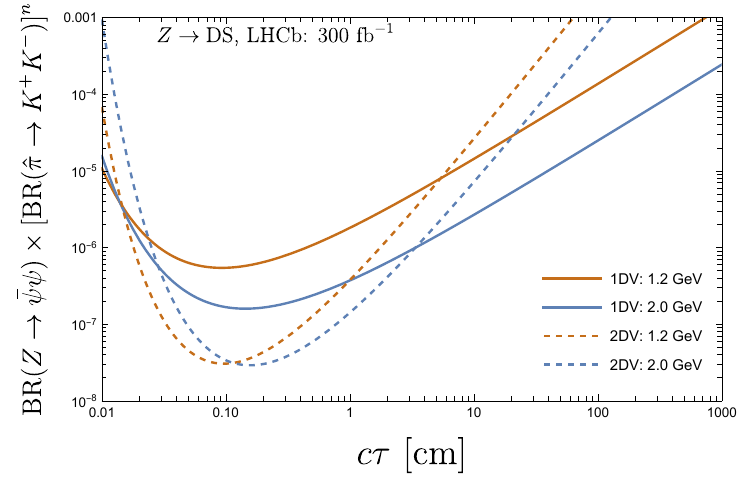}
\includegraphics[width=7 cm]{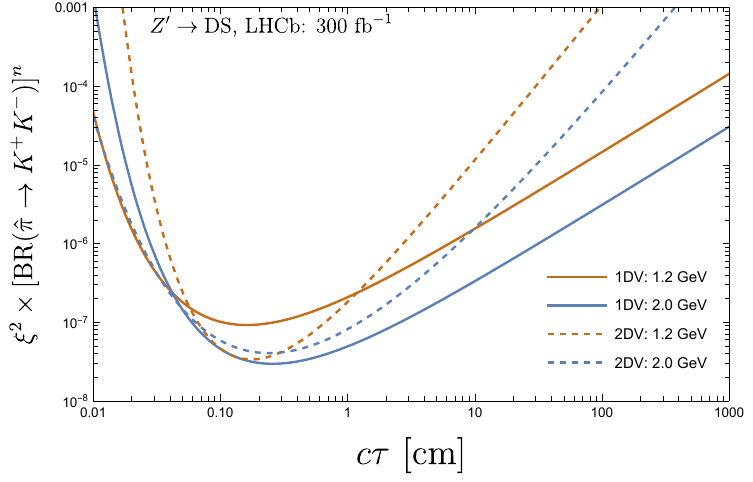}
\includegraphics[width=7 cm]{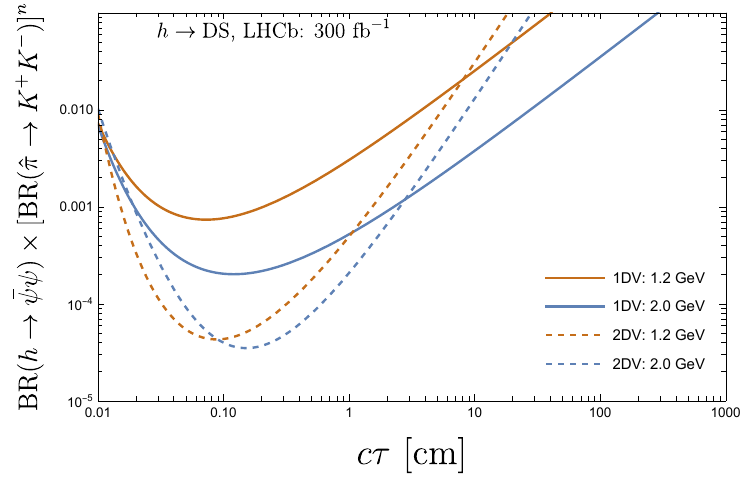}
\includegraphics[width=7 cm]{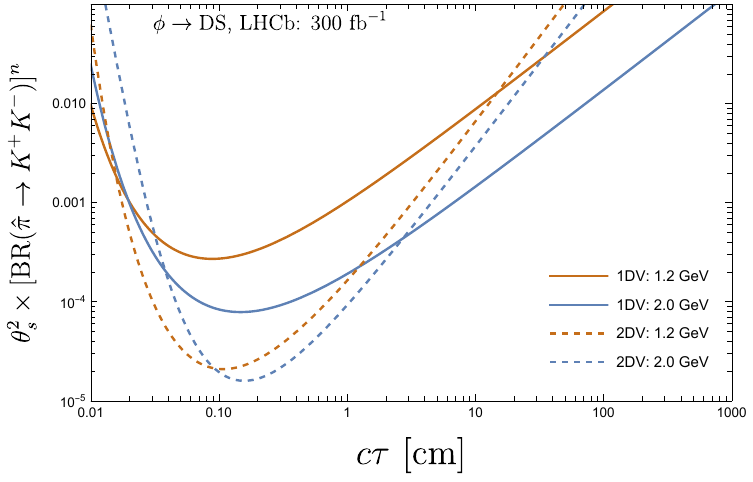}
\caption{The projected LHCb 95\% C.L. dark shower limits with $\hat{\pi} \to K^+K^-$ decays at the HL-LHC era. Here only $m_{\hat{\pi}}=1.2$ and 2~GeV benchmarks above 1~GeV are shown, while the background size largely drives the differences between 1DV limits. The 2DV limits from varying $m_{\hat{\pi}}$, on the contrary, are much closer.} \label{fig:LHCb_KK}
\end{figure}

We commence with the exclusive decays to $K^+K^-$, applicable to Higgs portal decays~\cite{Cheng:2021kjg,CidVidal:2019urm} or vector dark meson decays~\cite{Ilten:2018crw}. Dark shower signal samples are simulated using the same method as before, assuming $\hat{\pi}\to K^+K^-$ decays via the Higgs portal. We select benchmark dark pion masses to avoid overlap with the abundant $\phi$, $\Lambda^0$, and $D^0$ backgrounds. The selection rules in Ref.~\cite{CidVidal:2019urm} are applied at the truth level. Specifically, both $K^\pm$ tracks must satisfy $2<\eta<5$, $p_T>0.5$~GeV, and an impact parameter from the PV greater than $0.1$~mm. Furthermore, each reconstructed $KK$ DV must have $l_{xy}\in [6,25]$~mm, the longitudinal displacement $l_z<400$~mm, $p_T>10$~GeV, and an impact parameter from the PV less than $0.1$~mm. Two signal regions are defined by the DV's $l_{xy}$, namely $l_{xy}\in [6,10]$~mm and $l_{xy}\in [14,25]$~mm~\cite{CidVidal:2019urm}. Similar to previous analyses, no isolation criteria are imposed. A global detector efficiency factor of 0.7 is considered to account for detector effects.\footnote{The choice is more conservative than the efficiency adopted in Ref.~\cite{CidVidal:2019urm}.} We obtain the combinatorial background level from Ref.~\cite{CidVidal:2019urm}, which employs inclusive Pythia simulation. For non-isolated $KK$ DVs, the expected background in the range $m_{KK}\in[1,2]$~GeV is approximately $\mathcal{O}(10^5-10^6)$ when $L=15$~ab$^{-1}$, depending on the $l_{xy}$ bin. Such a high combinatorial level aligns with LHCb $D^0 \to K^+K^-$ studies~\cite{LHCb:2019dom,LHCb:2022vcc}. The substantial backgrounds diminish the search's sensitivity to signal events and render the result susceptible to systematic effects. By requiring two DVs with identical mass in a single event, the number of combinatorial backgrounds dramatically reduces by many orders of magnitude at the cost of lower signal yields~\cite{CidVidal:2019urm}. To define the 2DV signal region, two $K^+K^-$ DV satisfying all kinematic requirements above are required to have $l_{xy}\in [14,25]$~mm.\footnote{The requirement corresponds to the signal region $d_1$ in Ref.~\cite{CidVidal:2019urm}.} The result 2DV limits at the HL-LHC era are  plotted in Fig.~\ref{fig:LHCb_KK}. %

Besides the vertex detector, subdetectors several meters away from the interaction point could also be exploited to search for very long-lived LLPs with lifetimes longer than 100 ps. Decays before the magnet region can be reconstructed by the new Downstream algorithm~\cite{Gorkavenko:2023nbk}. Decays in the magnet region and the muon stations could also contribute. They are being studied by the LHCb collaboration~\cite{LHCb_LLP2024}. It would also be interesting to examine their reaches for the dark showers induced by the EW portal when more information is available.

\subsection{Dark Shower Sensitivity at Proposed Auxiliary Detectors}
\label{ssec:shower_aux}

\begin{table}[h!]
\centering
\begin{footnotesize}
\begin{tabular}{cccccc}
\hline
Detector & Geometry  & Displacement (m) & Volume (m) & Luminosity (fb$^{-1}$)\\
\hline
FASER~\cite{Ariga:2018uku}  & Cylinder & 0, 0, 480 & 0.2, 1.5 & 300\\ 
FASER 2~\cite{Ariga:2018uku} & Cylinder & 0, 0, 480 & 2, 5 & 3000\\
MATHUSLA(original)~\cite{MATHUSLA:2022sze} & Box & 75, 0, 118 & 30, 100, 100 & 3000\\
MATHUSLA(updated)~\cite{MATHUSLA_LLP2024}  & Box & 88.5, 0, 90 & 17, 40, 40 & 3000\\
Codex-b~\cite{Gligorov:2017nwh} & Box & 31, 2, 10 & 10, 10, 10 & 300\\
ANUBIS (Shaft)~\cite{Bauer:2019vqk} & Cylinder & 1.7, 51.5, 13.25 & 17.5,  57& 3000 \\
\hline
\end{tabular}
\end{footnotesize}
\caption{The simplified description of auxiliary detectors in this study. The column displacements describe the approximate distance of their geometric center from the closest LHC interaction point in 3 dimensions, with the last one always along the beam direction. The sizes of the detector volume are also presented. 
For detectors of the cylinder type, the sizes correspond to their diameter and length instead. The last column lists the corresponding integrated luminosity. For MATHUSLA, the original design of MATHUSLA may not be realized due to the funding limitation. A scale-back updated design is also listed. The original design of Codex-b also may not fully be used, but the new design is not available yet. For ANUBIS, the projection is based on the shaft-only configuration with lower background systematics.}
\label{tab:auxgeo}
\end{table}

Dark hadrons with proper decay lengths $\gtrsim \mathcal{O}$(m) will predominantly decay outside the multipurpose detector systems previously mentioned, resulting in reduced sensitivities which are apparent in, $e.g.$, Fig.~\ref{fig:Shower_CMS_1}. In such cases, LHC auxiliary detectors targeting LLPs can play a crucial role in dark shower searches due to their sensitivity to longer lifetime decays~\cite{Aielli:2019ivi,Curtin:2018mvb,Liebersbach:2024kzc}. 
Here we present limits for dark showers, which may be used generally.

Each auxiliary detector is simplified as an effective detector volume positioned away from the nearest LHC interaction point. Table~\ref{tab:auxgeo} provides a summarized description of these effective volumes, including their geometries, distances, and sizes.\footnote{We note the designs of some auxiliary detectors have yet to be frozen and may undergo future changes. However, all auxiliary detectors are sufficiently far from the closest interaction point, making the chance of an LLP decaying inside each part of the detector homogeneous. The constraints with alternative detector volumes are then well approximated by current ones rescaled by the volume ratios.} Only non-forward detectors, $i.e.$, ones with non-zero transverse displacements, are relevant for dark shower signals.  

\begin{figure}[h!]
\centering
\includegraphics[width=7 cm]{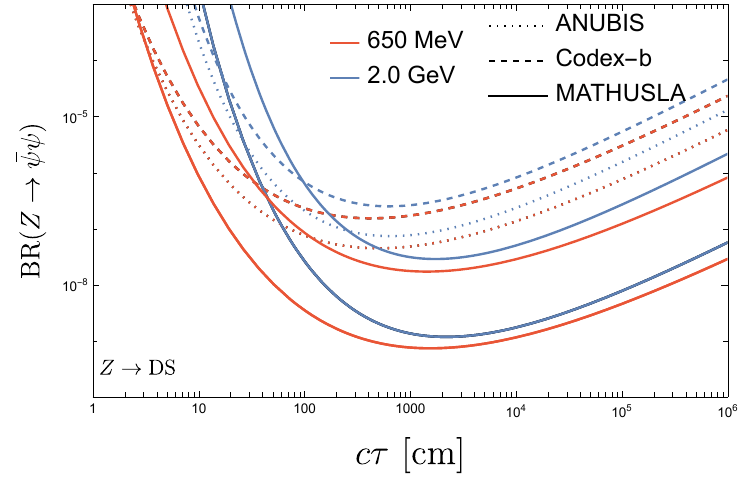}
\includegraphics[width=7 cm]{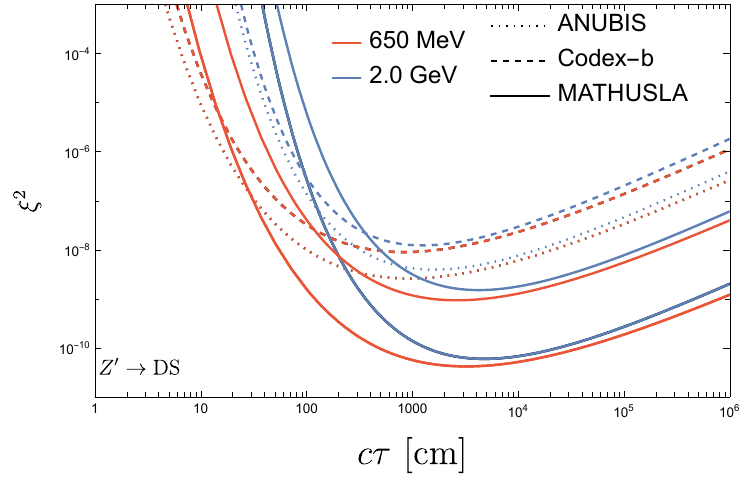}
\includegraphics[width=7 cm]{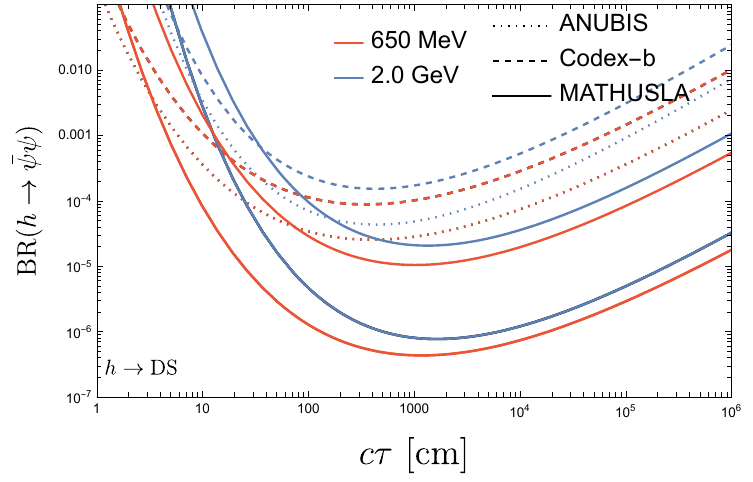}
\includegraphics[width=7 cm]{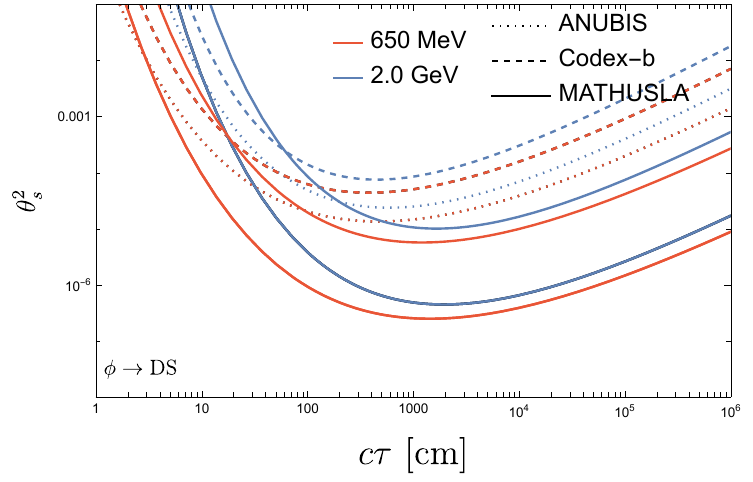}
\caption{The 95\% C.L. exclusion limits contours on displaced dark shower signals from the auxiliary detector ANUBIS, Codex-b, and MATHUSLA. Conventions are similar to Fig.~\ref{fig:Shower_CMS_1}. The two sets of MATHUSLA curves correspond to the original (stronger) and updated (weaker) designs, respectively.}
\label{fig:auxshower}
\end{figure}

The probability of each dark pion decaying within the effective detector volume is calculated based on its momentum and average lifetime. Since there is only limited shielding for the ANUBIS experiment, its SM background expectation is non-zero. Among the two ANUBIS configurations, $\mathcal{O}(10^5)$ background events are expected in the ceiling configuration, while in the shaft-only configuration, the backgrounds may be reduced to $\sim 10^3$~\cite{Satterthwaite:2839063}. Therefore, we focus on the shaft-only case and require a signal yield at ANUBIS $\sim 50$ to give the 95\% C.L. limit, following Ref.~\cite{Satterthwaite:2839063}. For all other searches, the majority of SM backgrounds vanish due to the presence of excessive shielding, making the background-free assumption legitimate. It is also assumed that all visible dark hadron decays can be accurately identified as the signal with an efficiency close to one. Consequently, the 95\% C.L. limits are obtained, corresponding to a signal yield $\sim 3$ according to Poisson statistics, as illustrated in Fig.~\ref{fig:auxshower}. All auxiliary detectors demonstrate strong discovery potential for dark hadrons with lifetimes exceeding the meter scale.  
For $m_{\hat{\pi}}\gtrsim 1$~GeV, the auxiliary detectors will benefit more as they are also sensitive to hadronic $\hat{\pi}$ decays. The $\hat{\pi} $ lifetimes needed for the optimal sensitivity is related to the distance of the auxiliary detector to the interaction point, as well as the dark pion boosts. Codex-b and ANUBIS are most sensitive to dark pions with $c\tau(\hat{\pi})$ between $\sim 3-10$~m. For MATHUSLA the ideal $c\tau(\hat{\pi})$ range is $\sim 10-50$~m. 

\subsection{Future High-Intensity Experiments}
\label{ssec:Darkshower_TeraZ}
Future lepton colliders often admit an operation phase running around the $Z$ pole with $\sqrt{s}\sim m_Z$ and very high luminosity~\cite{ILC:2013jhg,FCC:2018evy,CEPCStudyGroup:2018ghi,Bai:2021rdg}. At such future $Z$-factories, $\gtrsim \mathcal{O}(10^9)$ $Z$ bosons will be created on-shell with background processes highly suppressed. For circular $e^+e^-$ colliders, the typical projected number of produced $Z$ will exceed $10^{12}$, even larger than the total HL-LHC yield, ideal for studying exotic $Z$ decays to dark showers. Conversely, their $Z^\prime$, $h$, and $\phi$ yields are not as competitive.

\begin{figure}[h!]
\centering
\includegraphics[width=10 cm]{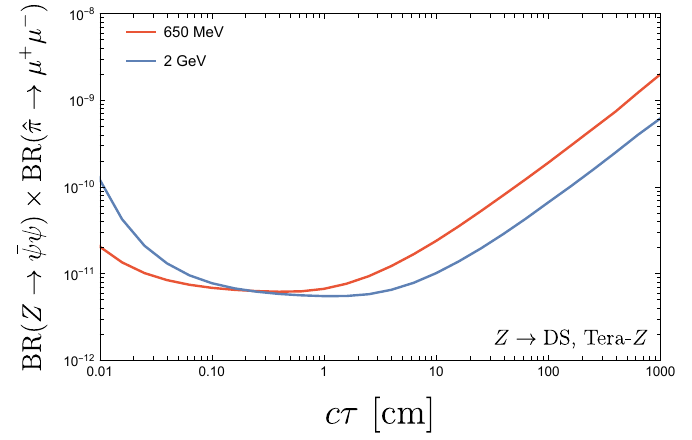}
\caption{The projected 95\% C.L. limits on  BR($Z\to \psi\bar{\psi}$)$\times$BR$(\hat{\pi}\to \mu^+\mu^-)$ at a Tera-$Z$ factory.}
\label{fig:Shower_TeraZ}
\end{figure}

The $e^+e^-\to Z \to$ dark shower signal samples are generated at the $Z$ pole ($\sqrt{s}=m_Z$). We make the following assumptions in our analysis. The dimuon DV signal is recognized if both $\mu$ satisfy $p_{T,\mu}>0.5$~GeV, $|p_{\mu}|>10$~GeV and $|\eta_\mu|<5$. In addition, the muon pair forms a DV and needs to have $p_{T,\mu\mu}>2$~GeV and $|\eta_\mu|<5$. A DV must have its invariant mass within the window of $[0.99~m_{\hat\pi},1.01m_{\hat\pi}]$ and its transverse displacement $\l_{xy} \in [0.5,100]$~cm. Finally, the signal yield is obtained by applying a global detector efficiency of 0.7 on each vertex as the benchmark value. Here we take the conservative assumption that the detector efficiency at future $Z$-factories will be no worse than its LHCb counterpart.
We also assume that the background is negligible for such well-reconstructed dimuon DV. Fig~\ref{fig:Shower_TeraZ} shows the projected 95\% C.L. exclusion limit on BR($Z\to \psi\bar{\psi}$)$\times$BR$(\hat{\pi}\to \mu^+\mu^-)$ at the Tera-$Z$, corresponding to $10^{12}$ on-shell $Z$ decays. The optimal limit is achieved when $c\tau(\hat{\pi})\in [0.1-10]$~cm. Within this optimal range, the signal efficiency is high enough, together with a very low background level expected, leading to a best reach of BR($Z\to \psi\bar{\psi}$)$\times$BR($\hat{\pi}\to \mu^+\mu^-$) close to the inverse of the total $Z$ number ($10^{-12}$).

\section{Limits on Dark Hadrons Produced by FCNC Meson Decays}
\label{sec:FCNC}

Here we study the accompanying flavor portal interactions in Eq.~\eqref{eq:FCNC4fermion}, which arise at one-loop EW processes. Light dark hadrons can thus be produced in SM heavy flavored meson FCNC decays besides dark shower. Since the process respects minimal flavor violation, the FCNC decay rates are generically too low to be observable for invisible dark sectors. However, when dark pions become LLP, many experiments will have better reach and become relevant. The rate mainly depends on $f_a$ and $m_{\hat{\pi}}$, accompanied by the logarithmic dependence on the theory's UV scale. From the effective interaction in Eq.~\eqref{eq:FCNC4fermion}, the exclusive $B$ FCNC decays producing dark pions are given as~\cite{Cheng:2024hvq}
\begin{equation}
\mathrm{BR}(B^{+,0} \to \{K^+ \hat{\pi}, K^{\ast 0} \hat{\pi} \} ) \approx \{ 0.92 , 1.1 \} \times 10^{-8}\, \bigg( \frac{1\;\mathrm{PeV}}{f_a} \bigg)^2 \bigg( \frac{\mathcal{K}_t}{10} \bigg)^2 \big\{ \lambda_{BK \hat{\pi}}^{1/2}, \lambda_{B K^\ast \hat{\pi}}^{3/2} \big\} \,,
\label{eq:FCNC1}
\end{equation}
where the dimensionless factor $\mathcal{K}_t$ depends on the UV cutoff logarithmically. The reference value $\mathcal{K}_t=10$ corresponds to a UV cutoff around TeV, depending on the details of the UV completion. The phase space factors $\lambda_{BK^{(\ast)}\hat{\pi}}\equiv\frac{4|p_{\hat{\pi}}|^2}{m_B^2}$ in the $B$ rest frame vary between 0.65 to $\sim$1 for the $m_{\hat{\pi}}$ range considered here. 

The inclusive FCNC decay widths, $e.g.$, $\Gamma(B\to X_s\hat{\pi})$ could be about an order magnitude larger than exclusive ones such as $\Gamma(B\to K^{(\ast)}\hat{\pi})$~\cite{Boiarska:2019jym}. Here $X_s$ stands for the inclusive hadronic final states that are strange-flavored. The inclusive FCNC decay rate can also be estimated from perturbative quark decay rates~\cite{Chivukula:1988gp}:
\begin{equation}
\label{eq:inclusive}
\Gamma_{(B\to X_s\hat{\pi})} \simeq \Gamma_{(b\to s \hat{\pi})}\frac{\Gamma_{(B\to X_c \ell \nu)}}{\Gamma_{(b\to c \ell \nu)}}~,
\end{equation} 
where $X_c$ is the charm-flavored inclusive final states.

For very light dark pions which can be produced in $K$ decay, its branching ratio can also be estimated as~\cite{Cheng:2024hvq}
\begin{equation}
\mathrm{BR}(K^{+} \to \pi^+ \hat{\pi}  ) \approx 3.9 \times 10^{-11}\, \bigg( \frac{1\;\mathrm{PeV}}{f_a} \bigg)^2 \bigg( \frac{\mathcal{K}_t}{10} \bigg)^2  \lambda_{K \pi \hat{\pi}}^{1/2} \,.
\end{equation}
The upper limits from NA62~\cite{NA62:2021zjw} on the branching ratio at 90\% CL are  $(3–6) \times10^{-11}$  for $m_{\hat{\pi}}$ in the range 0–110 MeV and $1 \times 10^{-11}$ for  $m_{\hat{\pi}}$ in the range 160–260 MeV, assuming $\hat{\pi}$ is invisible. Such a limit translates to a $f_a$ bound of $\mathcal{O}(1)$~PeV in these mass ranges. Later, we will see that the FCNC limits at LHC from $B$ meson decays have a similar or higher reach, covering a wider mass range. We thus focus on the $m_{\hat{\pi}} > 300$~MeV case and project LHC limits for FCNC $B$ decays.
 
 \subsection{FCNC Constraints and Projections at CMS and LHCb}
\label{ssec:FCNC_LHC}

The large statistics of heavy hadrons from $pp$ collisions enable LHC detectors to search for dark hadrons via FCNC interactions at one loop, as explicitly shown above. Many dark shower limits in Sec.~\ref{sec:DarkShower} are closely related to LLP searches from FCNC decays~\cite{CMS:2021sch,LHCb:2015nkv,LHCb:2020ysn}. While the analyses and backgrounds are highly analogous following the current framework, the kinematics of the signal are similar to heavy flavor decay products at the LHC, distinct from the dark shower case.

The $b$-hadron (here we focus on $B$ mesons) production and decay at the LHC are generated by Pythia8~\cite{Sjostrand:2014zea}, including all hard QCD processes. Their production rates in different $|\eta|$ ranges are corrected by matching with measured LHC fiducial rates~\cite{CMS:2016plw,LHCb:2020frr}. In all regions, we find Pythia 8 can provide reasonable predictions. The net $B$ meson yields from simulation are normalized by factors of 0.89 (0.76) in the central (forward) regions with $\eta\in [-2.5,2.5]([2,5])$, respectively. We also adopt the approximation that $B^0$ and $B^\pm$ FCNC decay rates are identical.

Since the FCNC decay rates to dark hadrons are suppressed by $f_a$, 
off-diagonal CKM matrix elements, and loop factors, we expect no more than one dark pion to be produced per FCNC decay.\footnote{Since the dark QCD could also shower after the dark fermion pair is produced, it is also possible that more than one dark hadron are created by the FCNC decay, $e.g.$, $B\to K+n\hat{\pi}$ process. In this work, we only consider the case of single dark pion production.} The signal feature at high-energy colliders will be one DV produced along the direction of a heavy-flavor initiated jet. In this case, the accompanying SM particle like $K^0$ or $\pi^0$ from FCNC decay may not be reconstructed. Meanwhile particles from the rest of the jet or meson decays make the dimuon DV unisolated.\footnote{ When the dark hadron is light enough that it is still relativistic in the $B$ meson's rest frame, we expect the kinematics from $B\to X_s\hat{\pi}$ to be similar to the exclusive $B\to K^{(\ast)}\hat{\pi}$ decays. Consequently, our strategy is potentially applicable to inclusive FCNC decays.}

\begin{figure}
\centering
\includegraphics[width=10 cm]{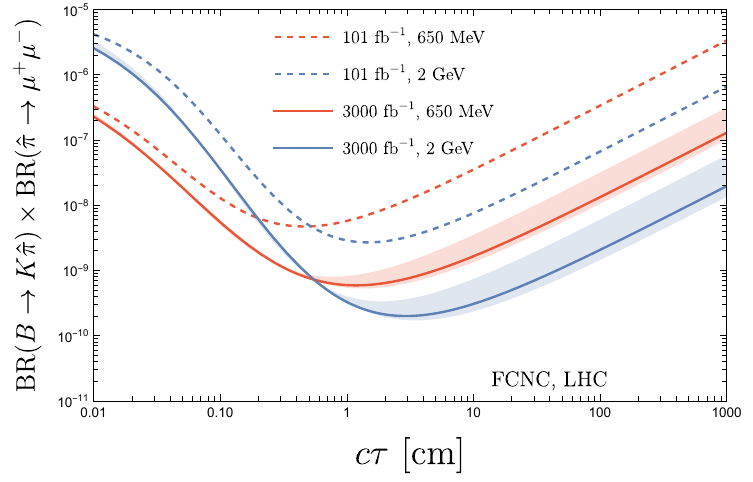}
\caption{Current and projected (HL-)LHC  95\% C.L. limits on dark hadrons from FCNC $b$-hadron decays based on the CMS dimuon DV scouting strategy. The solid curves and the corresponding band stands for the HL-LHC projections and uncertainties, while the current limit is shown as dashed curves.}
\label{fig:FCNC_main}
\end{figure}
The FCNC limit from the CMS dimuon scouting is obtained following the same analysis in Sec.~\ref{ssec:darkshower_main}, counting for 1DV signal with a fixed amount of background. The projected limits for exclusive $B\to K\hat{\pi}(\to \mu^+\mu^-)$ decays are shown in Fig.~\ref{fig:FCNC_main}. Both current LHC and projected HL-LHC limits come with similar features found in dark shower searches.\footnote{Due to the finite displacement of $B$ decays, the sensitivity for short dark pion lifetime case ($\lesssim 10^{-3}$~cm) will asymptotically converge to a fixed value  $\lesssim 10^{-5}$.} The available CMS data can constrain the exclusive FCNC decay rate down to $\mathcal{O}(10^{-9}-10^{-8})$ level. The limit will further strengthen by about an order of magnitude in the high luminosity era.

\begin{figure}
\centering
\includegraphics[width=10 cm]{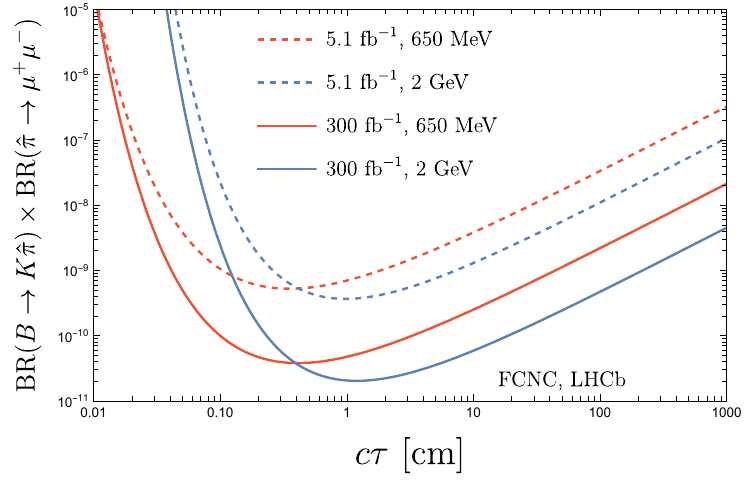}
\caption{The 95\% C.L. $B$ meson FCNC decay to dark hadron limits given by LHCb. Dashed curves are current ones.}
\label{fig:FCNC_LHCb}
\end{figure}

The exclusive upper bounds on BR($B\to K \hat{\pi}$) derived from LHCb also follow the approaches detailed in Sec.~\ref{ssec:shower_LHCb}. In contrast to the dark shower search, the dark pion will have a small extra displacement due to the finite $B$ lifetime. We thus apply the ``inclusive" type of background in Ref.~\cite{LHCb:2020ysn}, in which the DV momenta do not have to be aligned with their displacements from the PV. The corresponding backgrounds increase by $\mathcal{O}(10^2)$ times higher than the ``prompt" type, depending on $m_{\mu\mu}$ and $p_{T,\mu\mu}$. However, such ``inclusive" backgrounds are still significantly smaller than their CMS counterparts. The current and projected limits are plotted in Fig.~\ref{fig:FCNC_LHCb}. Compared to their CMS counterparts, the LHCb constraints are more stringent due to the substantial $b\bar{b}$ production in the forward region and lower background. The FCNC decay sensitivity obtained this way is comparable with the one from exclusive searches where the $B$ meson is fully reconstructed~\cite{LHCb:2015nkv,LHCb:2016awg}. Due to different search strategies, the method of Fig.~\ref{fig:FCNC_LHCb} performs better when the dark pion decay length is $\gtrsim 1$~mm, more suitable for probing the larger $f_a$ regime. Conversely, the search in Refs.~\cite{LHCb:2015nkv,LHCb:2016awg} does not place explicit requirements on the DV's $l_{xy}$, making the search more capable of short-lifetime or even prompt dark pions.

\subsection{FCNC Constraints at Auxiliary Detectors at HL-LHC}
\label{ssec:FCNC_aux}
The heavy-flavored mesons are efficiently produced at LHC in the central and forward regions, making most auxiliary detectors sensitive to FCNC signals. The same Pythia8-generated FCNC signal samples are used for detectors proposed to probe the central region. The FORESEE~\cite{Kling:2021fwx} package is applied to derive corresponding projected sensitivities to account for signal yields in the highly forward region. 
Similar to the approach in Sec.~\ref{ssec:shower_aux}, the 95\% C.L. exclusion limits are represented by the signal yield of 3 that falls within the effective volume, as listed in Table~\ref{tab:auxgeo}, except for ANUBIS, where 50 signal events are required.

\begin{figure}[h!]
\centering
\includegraphics[width=12 cm]{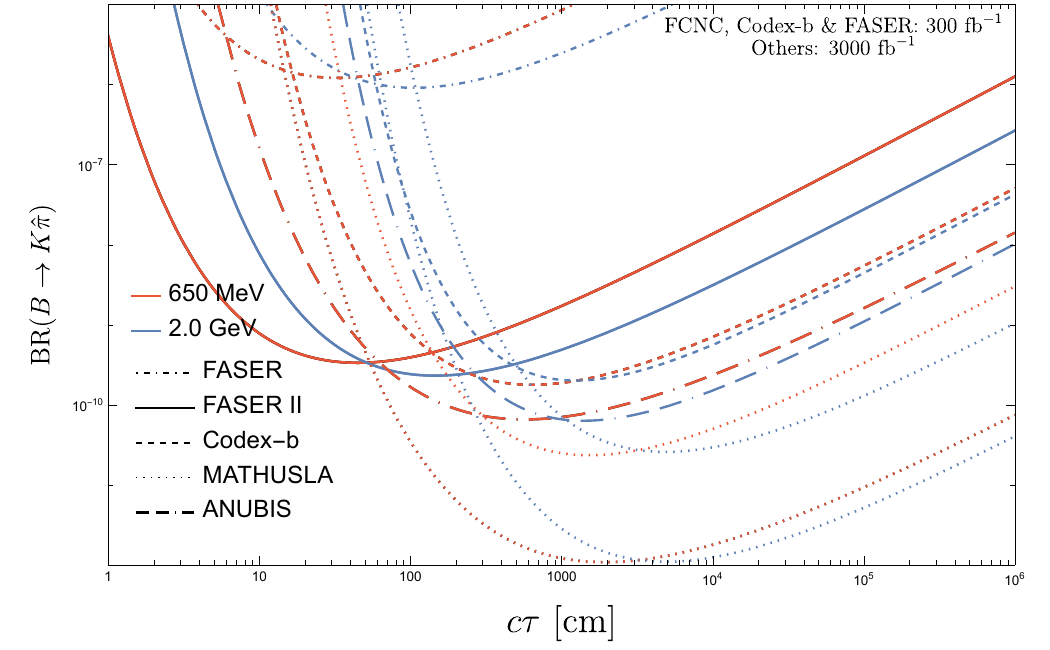}
\caption{The 95\% C.L. limits on exclusive BR($B\to \hat{\pi} K$) at various auxiliary detectors. The two sets of MATHUSLA curves correspond to the original and updated designs respectively.}
\label{fig:auxFCNC}
\end{figure}

The results for various auxiliary detectors are shown in Fig.~\ref{fig:auxFCNC}. The optimal performance of each detector largely depends on the geometric properties of their effective volume. MATHUSLA, as the largest detector concerned here, can probe an FCNC branching ratio down to $\mathcal{O}(10^{-11})$ level for a wide range of $m_{\hat{\pi}}$. Such optimal limit is obtained when $c\tau(\hat{\pi})\in [10,100]$~m, similar to the dark shower case discussed in Sec.~\ref{ssec:shower_aux}. The limits are followed by ANUBIS and Codex-b ones, which are about a few times weaker with shorter optimal $c\tau(\hat{\pi})$. Meanwhile, the ample $b$ production in the forward region benefits detectors in the forward region. While the FASER limits are not as competitive due to its small volume, the FASER2 constraints are comparable to Codex-b ones with slightly weaker optimal reaches in BR($B\to K\hat{\pi} $). The forward detectors are most sensitive when $c\tau(\hat{\pi})\in [10,100]$~cm, significantly shorter than the other auxiliary detectors due to the large $B$ meson boost in the highly-forward region.

\subsection{FCNC Constraints from Other Experiments}
\label{ssec:FCNC_intensity}

\textbf{CHARM:}
Experiments with lower energy scales, either fixed-target or beam dump type, benefit from high-intensity beam lines~\cite{Bross:1989mp,NA482:2015wmo,MicroBooNE:2016pwy,NA62:2017rwk,Berlin:2018pwi,T2K:2019bbb}. However, constraints from such experiments rely on detailed knowledge of the beam's interaction with the material, especially various nuclei, making predictions more challenging.

\begin{figure}[t!]
\centering
\includegraphics[width=13cm]{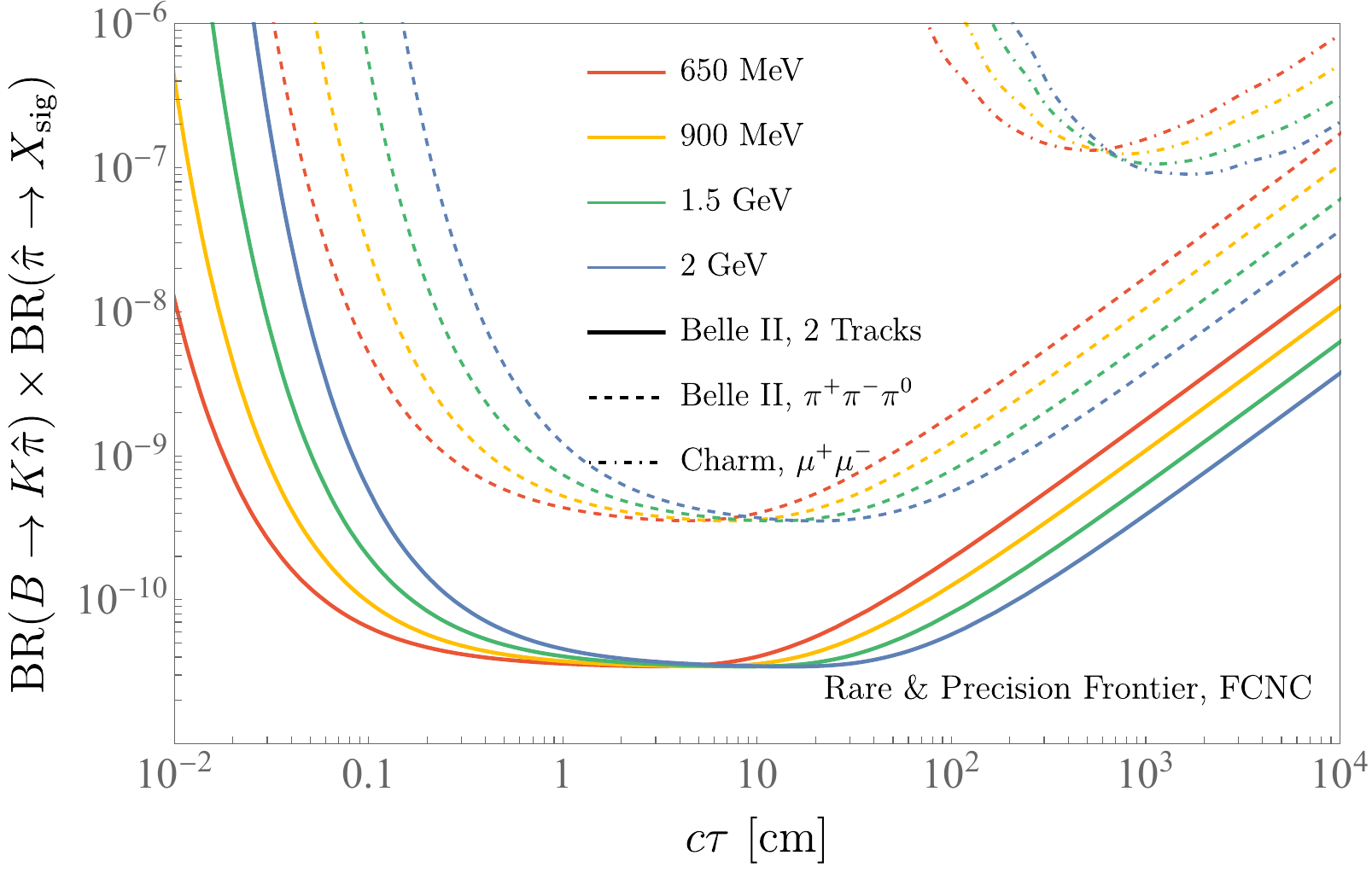}
\caption{The 95\% C.L. exclusion limits ($N_{\rm sig}=3$ contours) at Belle II (50 ab$^{-1}$) and Charm, where the signal final state $X_{\rm sig}$ is created by dark hadron $\hat{\pi}$ exclusive $B\to K+\hat{\pi}$ decays.} 
\label{fig:BelleII_1}
\end{figure}

For bounds of this type, we focus on the CHARM experiment~\cite{CHARM:1985anb}. We follow the recast in Ref.~\cite{NA62:2023nhs} using muon pair final states.\footnote{We thank Reuven Balkin for pointing out the proper reference here.} 
 For different dark hadron mass benchmarks, the results are shown in Fig.~\ref{fig:BelleII_1}, with the maximum sensitivity reached for dark hadron lifetimes of $\sim \mathcal{O}(10)$~m.

\noindent\textbf{Belle II:}
Lepton collider experiments operating at $\sqrt{s}\simeq 10$~GeV are also known as $B$ factories as $\Upsilon$(4S) can be produced on resonance, which will promptly decay to two $B$ mesons with large cross section.\footnote{One may also consider dark shower signals at $B$ factories. However, in our EW portal benchmarks, the mediating bosons are too heavy to be produced on-shell at $B$ factories and leading to suppressed signal yields. We found no particular advantage of dark shower search at $B$ factories compared to the LHC. For dark shower searches at Belle II induced by other portal interactions, see Ref.~\cite{Bernreuther:2022jlj}.} The leading $B$ factory for the next decades is Belle II~\cite{Kou:2018nap}. For the projection at Belle II, we adopt the method in Ref.~\cite{Filimonova:2019tuy}. The $B$ mesons from $\Upsilon(4S)$ decays have a fixed boost of $\beta\gamma\simeq 0.284$ along the beam axis, while their transverse boosts are negligible. To reconstruct the DV, both tracks need to leave enough hits to be reconstructed and form DVs~\cite{Bernreuther:2022jlj}. The signal efficiency and the effective volume of the Belle II detector become smaller than the detector's geometric size, with negligible backgrounds~\cite{Ferber:2022ewf}. Here, we adopt a simplified detector geometry of the Belle II vertex detector and central drift chamber~\cite{Kou:2018nap,Ferber:2022ewf}. All dark pions that decay inside or before reaching the vertex detector or the central drift chamber with $60>l_{xy}>0.2$~cm are kept. The macroscopic $l_{xy}$ and a narrow resonance peak can reduce combinatorial backgrounds to a negligible level for final states with only two charged tracks, i.e., $\mu^+\mu^-$, $\pi^+\pi^-$, or $K^+K^-$.\footnote{For electron pair final states, further cuts are needed to remove large backgrounds from converted photons.}  Assuming a signal efficiency of $\sim 1$ and negligible backgrounds, the 95\% C.L. exclusion limit then overlaps with the $N_{\rm sig}=3$ contour in Fig.~\ref{fig:BelleII_1}. The experiment could probe an exclusive FCNC $B\to K+\hat{\pi}$ decay with a branching ratio down to $\mathcal{O}(10^{-10})$ level when $c\tau(\hat{\pi})$ is between 0.1-10~cm.

Besides final states with only charged particles, $B$ factories with clean collision environments may also provide opportunities to probe final states with neutral components such as $\gamma$ or $\pi^0$. The $\pi^0 \to\gamma\gamma$ from $B$ decays can be identified from the diphoton mass peak with low backgrounds, with the reconstruction efficiency varying between $20\%$ to $60\%$~\cite{Stengel:2019vma}. The two photons and tracks can also reconstruct the $B$ resonance with a relative mass resolution of a few percent~\cite {Aritra:3255}, further reducing the combinatorial background. Taking the $\pi^+\pi^-\pi^0$ final state as a benchmark, the number of $\pi^+\pi^-$ decay vertices in the effective volume is traced. In this case, the minimum $l_{xy}$ requirement increases to 0.9~cm to eliminate remaining $c$-hadron backgrounds. The signal efficiency is also reduced by a factor of 0.2, representing the loss from $\pi^0$ and $B$ resonance reconstruction. Finally, we only consider $B^\pm\to K^\pm \hat{\pi}(\to \pi^+\pi^-\pi^0)$ decays, given the $K^\pm$ track can be used for the $B$ resonance reconstruction with high efficiency. Assuming all the above techniques make the backgrounds negligible, we plot the 95\% C.L. limit at Belle II in Fig.~\ref{fig:BelleII_1}. We expect such a limit, although not as strong as those with only two tracks in terms of BR values, will be complimentary when searching for dark pions heavier than 1~GeV due to their high BR($\hat{\pi}\to \pi^+\pi^-\pi^0$).

\begin{figure}[h!]
\centering
\includegraphics[width=10 cm]{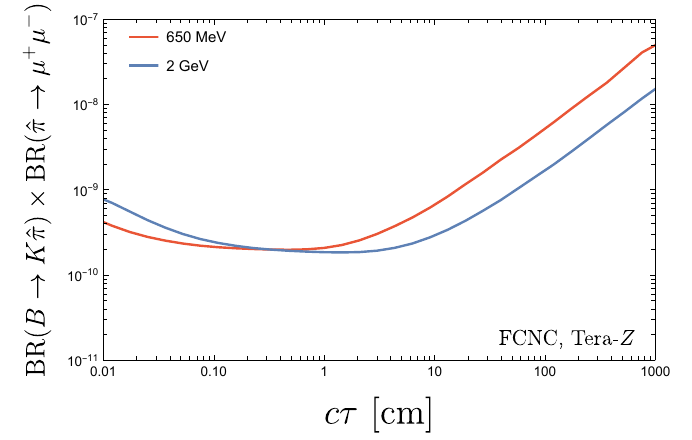}
\caption{The 95\% C.L. FCNC $B\to K(\hat{\pi}\to\mu^+\mu^-)$ limits from Tera-$Z$.}
\label{fig:FCNC_TeraZ}
\end{figure}

\noindent\textbf{$\bm{Z}$ Factories}
The future $Z$ factories will also be sensitive to FCNC meson decays as many flavored hadrons will be produced from $Z\to b\bar{b},~c\bar{c}$, and $s\bar{s}$ decays. Here we focus on $B$ mesons from $Z\to b\bar{b}$ decay. 
The signal region definition from dimuon DVs and the background assumption are the same as described in Sec.~\ref{ssec:Darkshower_TeraZ}, assuming $10^{12}$ $Z$ bosons produced on-shell, with the results shown in Fig.~\ref{fig:FCNC_TeraZ}.

\section{Combined Constraints on EW-portal Benchmark Models}
\label{sec:model}

With the general bounds obtained in the previous sections, one can derive model-specific bounds on the model parameters. Given a specific model, one can calculate the branching ratios of $Z$ and $h$ decays to the dark sector, the production cross sections of $Z'$ and $\phi$, the effective ALP decay constant $f_a$ of the dark pions, and the FCNC $B$ decay rates. The reaches in the model parameter space from various experiments can then be compared and combined directly. In this section, we will illustrate this with a couple of EW-portal benchmarks discussed in Refs.~\cite{Cheng:2019yai,Cheng:2024hvq}. The first one is a $Z$-portal model where light dark quarks acquire small couplings to $Z$ by mixing with heavy EW doublet fermions through the Higgs VEV, as shown in the middle panel of Fig.~\ref{fig:mixing_diagrams}. In the second benchmark, the couplings arise from the mixing between $Z$ and a dark $Z'$, which couple to the light dark quarks. The Higgs-initiated dark showers are also present in both models, and the scalar $\phi$ can exist in the second model, but their contributions are subleading. The lifetimes of dark pions are typically too long for collider experiments in pure Higgs-portal models unless additional ingredients are added to enhance the dark pion decays. Therefore, we focus on $Z$- and $Z'$-initiated dark showers together with FCNC $B$ decays. However, we emphasize that they only serve as examples but do not necessarily represent general results due to the large parameter space to be explored. For concreteness, we take the number of dark colors $N_d=3$ and the dark flavor number $N=2$, with $N^2-1=3$ light dark pions in both cases.
For the heavy doublet fermion mixing model~\cite{Cheng:2021kjg}, the light dark quarks $\psi_L$, $\psi_R$ couple to some heavy EW doublets $Q_L$, $Q_R$,
\begin{equation} \label{eq:LUV}
- \mathcal{L}_{\rm UV} = \overline{Q}_L  \byuk \psi_R  {H} + \overline{Q}_R \byukt\psi_L {H} + \overline{Q}_L  \bmass Q_R +  \overline{\psi}_L \bomega \psi_R + \text{h.c.}\,,
\end{equation}
where $\byuk$, $\byukt$, $\bmass$, and $\bomega$ are $N\times N$ ($N=2$) matrices in flavor space. We assume $\byukt=0$ for this study.\footnote{The Higgs portal can become dominant if $\byukt \sim \byuk$, but in this case their product is constrained by the exotic Higgs branching ratio, and the lifetimes of the dark pions are typically too long for collider searches.} The light dark quark mass eigenstates $\psi'$ couple to $Z$ through the mixing $Y v/(\sqrt{2} M)$ with the heavy doublets $Q$. The branching ratio of the $Z$ decay to dark fermions is\footnote{This formula assumes that the heavy fermions have a universal mass $M$. More general results are shown in Ref.~\cite{Cheng:2021kjg}.}
\begin{equation}\label{eq:Z_BR}
\text{BR} (Z \to \psi' \overline{\psi}^{\prime} ) \approx  1.8 \times 10^{-4} \, \bigg(\frac{ N_d \text{Tr} (  {\byuk}\byuk^\dag {\byuk}  \byuk^\dag ) }{3} \bigg) \left( \frac{1\, \text{TeV}}{M}\right)^4 .
\end{equation}
Assuming $CP$ conservation in the dark sector, the decays of $CP$-odd dark pions $\hat{\pi}^{1,3}$ to SM particles through the $Z$ portal can be described by an effective ALP interaction. For simplicity, we take a benchmark where $\hat{\pi}^{1,3}$ have the same $f_a$ and thus have the same lifetime, while the $CP$-even $\hat{\pi}^2$ has a long lifetime and is effectively invisible. The details of the benchmark choice are given in App~\ref{app:benchmark}. The effective ALP decay constant for $\hat{\pi}^{1,3}$ is parametrically given by
\begin{equation}
f_a^{1,3} \approx 5.7\;\mathrm{PeV} \, \bigg( \frac{1\;\mathrm{GeV}}{f_{\hat{\pi}}} \bigg) \bigg( \frac{M}{1~\text{TeV}}\bigg)^2  \bigg( \frac{1}{\text{Tr}[\byuk^\dagger \byuk]} \bigg) \sim \frac{M^2}{\text{Tr}[\byuk^\dagger \byuk] f_{\hat{\pi}}}~,
\end{equation}
where $f_{\hat{\pi}}$ is dark pion decay constant in dark QCD defined analogously to $f_\pi$ in QCD.

In the $Z-$dark $Z'$ mixing model~\cite{Cheng:2024hvq}, the light dark quarks $\psi_{Li}$, $\psi_{Ri}$ couple to the (massive) dark $Z'$ with charges $x_{Li}$, $x_{Ri}$, and an overall coupling $g_D$. The dark $Z'$ mixes with SM neutral gauge bosons through both kinetic and mass mixings,
\begin{equation}
\mathcal {L}_{\rm mix} = -\frac{\sin \chi}{2} \hat{Z}'_{\mu\nu} \hat{B}^{\mu\nu} + \delta \hat{M}^2 \hat{Z}^{\prime \mu} \hat{Z}_\mu ,  \label{eq:Zhat2}
\end{equation}
where $\hat{B}^{\mu\nu}$ is the hypercharge field strength, and all ``hatted'' quantities stand for ones before mixing. The mass mixing is necessary for $CP$-odd dark pion decays, as the decays only go through the longitudinal mode. 
The overall mixing $\xi$ (which is a combination of the mass and kinetic mixings) between $Z$ and $Z'$ determines the branching ratio of $Z$ decay into dark quarks,
\begin{equation}
\text{BR} (Z \to \psi' \overline{\psi}^{\prime} ) \approx  1.4 \times 10^{-4} \, \bigg(\frac{N_d}{3}\bigg) \bigg( \frac{\xi}{0.01}\bigg)^2 g_D^2 \sum_i (x_{Li}^2+x_{Ri}^2).
\end{equation}
and also the production cross section of the $Z'$ boson.\footnote{$\xi$ takes the simple form of $\delta\hat{M}^2/(m_Z^2-m_{Z^\prime }^2)$ in the limit of $1\gg\delta \hat{M}^2/m_Z^2\gg \chi$. } (The $Z'$ decay to dark quarks is close to 100\%, assuming it doesn't couple to SM states directly.) We refer the reader to Ref.~\cite{Cheng:2024hvq} for a detailed description of the model and the relation between $\xi$ and $\chi$, $\delta \hat{M}^2$. Similar to the previous case, we take a benchmark in which $\hat{\pi}^{1,3}$ have the same $f_a$ and $\hat{\pi}^2$ is effectively stable at the colliders. The benchmark parameters are described in App~\ref{app:benchmark}. The effective ALP decay constant is parametrically given by 
\begin{equation}
f_a^{1,3}\approx 3.9\;\mathrm{PeV} \, \left( \frac{1\;\mathrm{GeV}}{f_{\hat{\pi}}} \right) \bigg( \frac{10^{-2}}{\delta \hat{M}^2/ \hat{M}_{Z}^2}\bigg)  \left( \frac{ M_{Z^\prime}}{60 \text{ GeV}} \right)^2\, \sim \frac{M_Z^2  M_{Z'}^2 }{ g_D g_Z \delta M^2 f_{\hat{\pi}} },
\end{equation}
where $g_Z =\sqrt{g^2+g^{\prime 2}}$ is the $Z$ coupling.

The FCNC production of the dark pions mainly depends on the effective ALP decay constant $f_a$, so it is insensitive to model details (except for some logarithmic dependence on the UV completion scale) as long as it gives the same $f_a$. Therefore, it provides an independent probe for the dark shower search. Since the search strategy is based on single DV without isolation requirements, the SM decay products of $B$ have no direct impact on the analysis. It is thus reasonable to assume the signal efficiency of the inclusive $B\to \hat{\pi}+X_s$ decays will be similar to the exclusive $B\to \hat{\pi}+K^{(\ast)}$ decay ones. According to Eq.~\eqref{eq:inclusive}, the inclusive process has a decay rate $\sim$ 7 times larger than the exclusive ones discussed. To better demonstrate the potential of the flavor probes, we will plot the inclusive $B$ decay constraints based on the assumption above. The relative strengths in experimental bounds between the dark shower and FCNC productions can only be evaluated within specific models, which is the task of this section.

\begin{figure}[h!]
\centering
\includegraphics[width=7.5 cm]{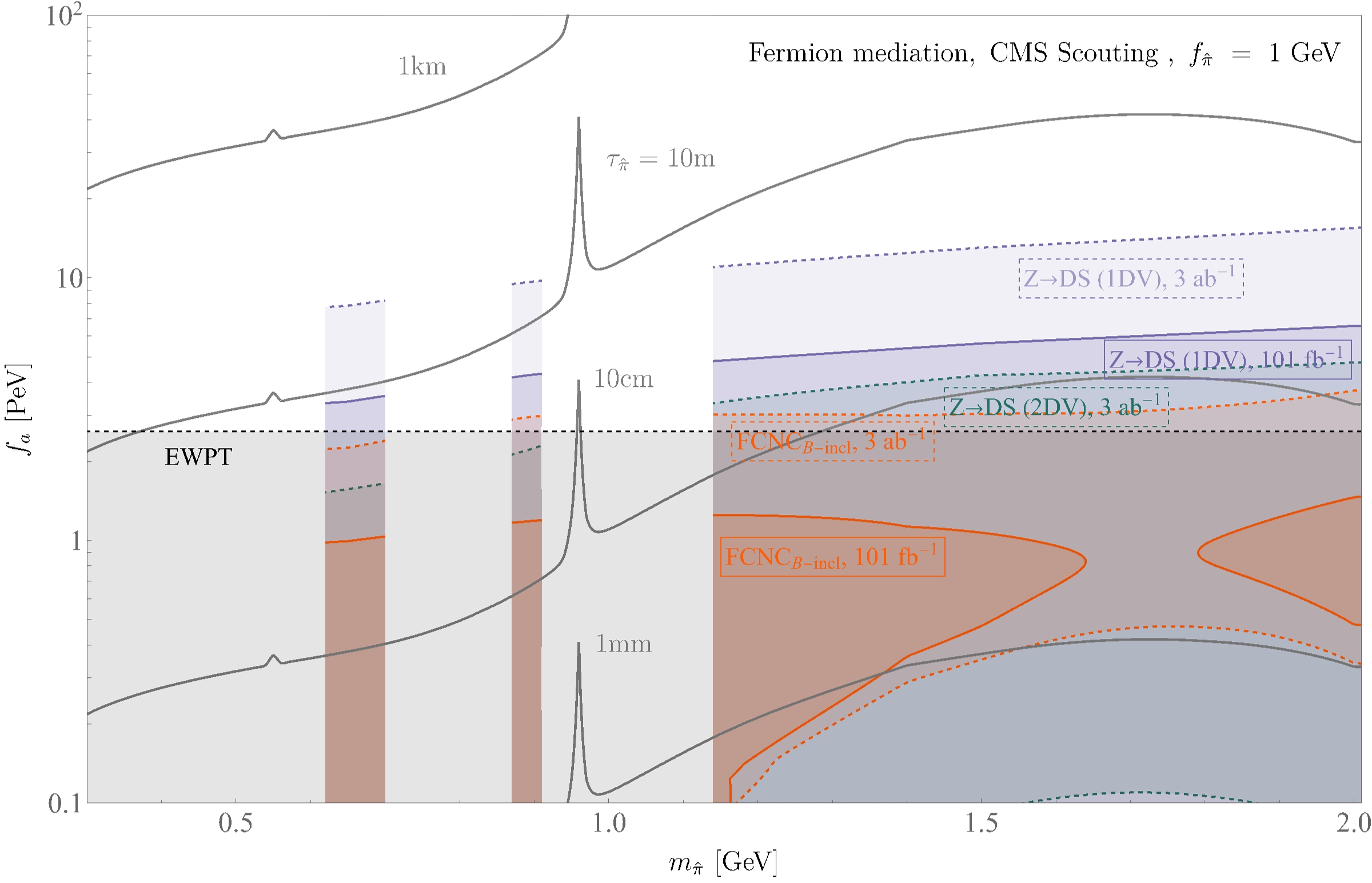}
\includegraphics[width=7.5 cm]{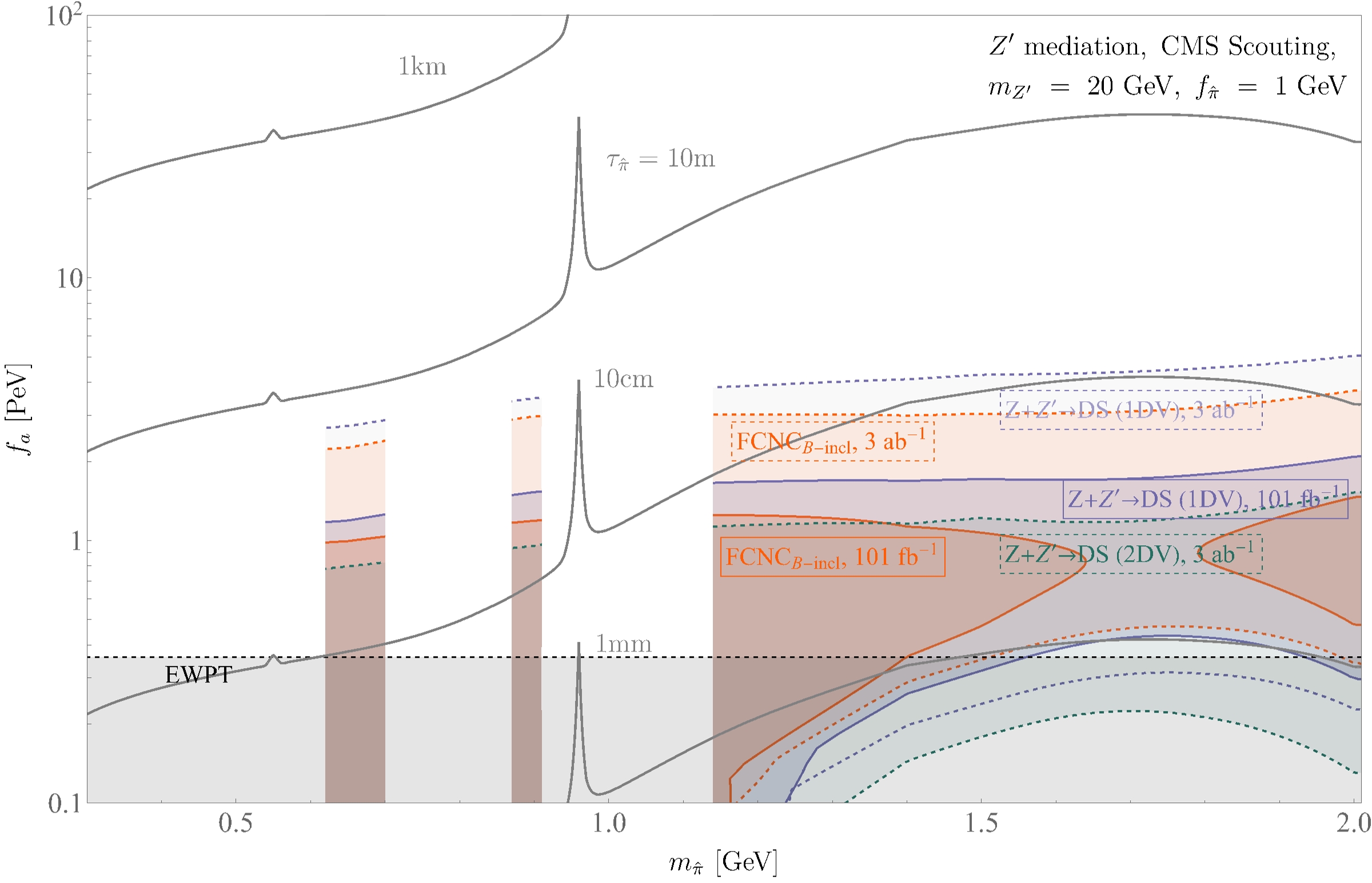}
\caption{Model-specific 95\% C.L. exclusion limits from the dimuon DV scouting strategy at (HL-)LHC with various channels. \textbf{LEFT: } Exclusion limits for a fermion doublet model. \textbf{RIGHT: } Exclusion limits for a $Z^\prime$ model with $m_{Z^\prime}$=20~GeV.}
\label{fig:modeldependent_1}
\end{figure}

\noindent \textbf{Constraints from CMS} In Fig.~\ref{fig:modeldependent_1}, we plot CMS scouting search limits for the two benchmark models mentioned above (heavy doublet fermion mixing and dark $Z'$) with benchmark parameters given in App~\ref{app:benchmark}. In both panels, we display the limits in the $m_{\hat{\pi}}-f_a$ plane, with $f_{\hat{\pi}}$ fixed to be 1 GeV.  The dark pion lifetime is entirely determined by $f_a$ and $m_{\hat{\pi}}$ as given in Eqs.~\eqref{eq:decaywidth}, \eqref{eq:decaylength} and shown as the light gray contours. The current EWPT constraints are also shown as shaded regions independent of $f_a$. We adopt fit values from Ref.~\cite{Cheng:2024hvq} for the dark $Z^\prime$ mediation case. For heavy fermion mediators, the calculations are detailed in App.~\ref{app:benchmark}.\footnote{Since $f_a$ and the oblique parameters have different $M$ and $\byuk$ dependence, the constraint differs with parameter choice. Here, we adopt a nominal lower limit of $M\gtrsim 1$~TeV. For higher $M$ values, the EWPT lower bound on on $f_a$ scales proportional to $M$.}

For the dark showers, the current limits and the HL-LHC projections of the 1DV dimuon scouting search are shown. For the dark $Z'$ model, the contributions from $Z$ and $Z'$ are summed, with the $Z'$ contribution dominating due to the large overall production rate for a light $Z'$. On the other hand, a lighter $Z'$ also lowers the value of $f_a$. The overall result is that the same signal rate corresponds to a higher $f_a$ value in the heavy doublet fermion mixing model where only dark showers from $Z$ decay are included. The current limits have surpassed the EWPT constraints of these benchmark models. The HL-LHC could improve the limits on $f_a$ by a factor of $\sim$2. We also plotted the future projections of the 2DV limits from the dark showers. They tend to constrain the lower $f_a$ region due to their limited sensitivity for longer lifetimes. As we discussed earlier, the projections are highly conservative, so further improvements are possible. The multi-DV events would provide a direct indication of the dark shower if discovered.

The $B$ FCNC inclusive decay limits from the scouting search are also shown in both panels of Fig.~\ref{fig:modeldependent_1} according to the general prediction of FCNC decay rates as functions of $f_a$ and $m_{\hat{\pi}}$. They are scaled up from the exclusive limits of Sec.~\ref{sec:FCNC} using Eq.~\eqref{eq:inclusive}. In both heavy doublet fermion mixing  and dark $Z^\prime$ models, the corresponding FCNC limits are identical if the benchmark models have the same $\mathcal{K}_t$ and the same decaying dark pions. Such predictions are even independent of $f_{\hat{\pi}}$ as it is included in the definition of $f_a$. Therefore, the limits given are also applicable to general ALPs with the same couplings. The relative strengths of the dark shower constraints and FCNC constraints depend on $f_{\hat{\pi}}$. For $f_{\hat{\pi}}=1$~GeV adopted in the plots, the dark showers give stronger constraints. For a fixed $f_a$, increasing $f_{\hat{\pi}}$ implies smaller couplings of $Z$ to the dark quarks and smaller mixing between $Z$ and $Z'$. This will reduce the production rate of the dark showers and hence their search reaches, while the FCNC constraints stay more or less unchanged. Therefore, for large enough $f_{\hat{\pi}}$, a stronger limit may come from FCNC instead.

\begin{figure}[h!]
\centering
\includegraphics[width=7.5 cm]{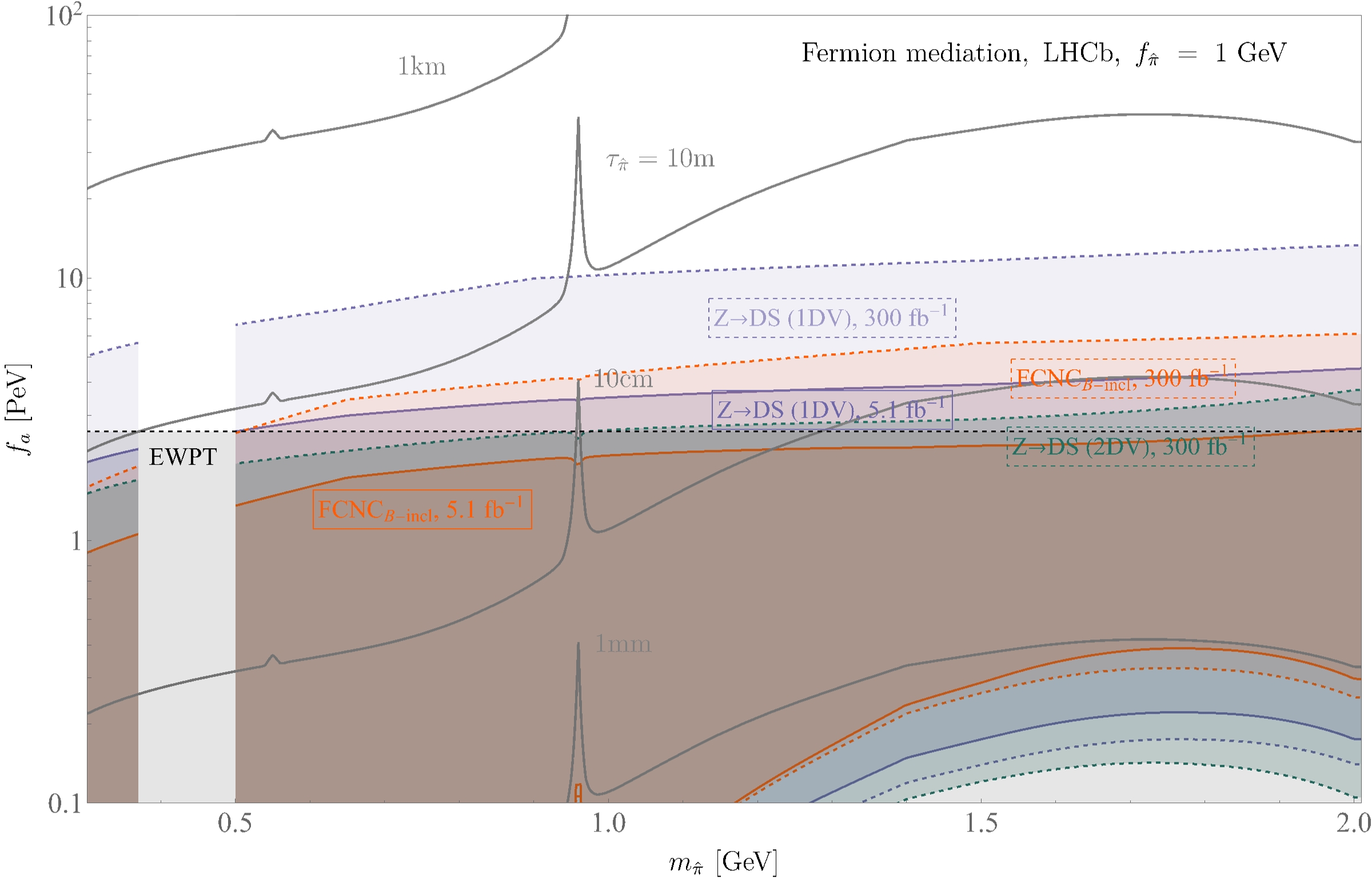}
\includegraphics[width=7.5 cm]{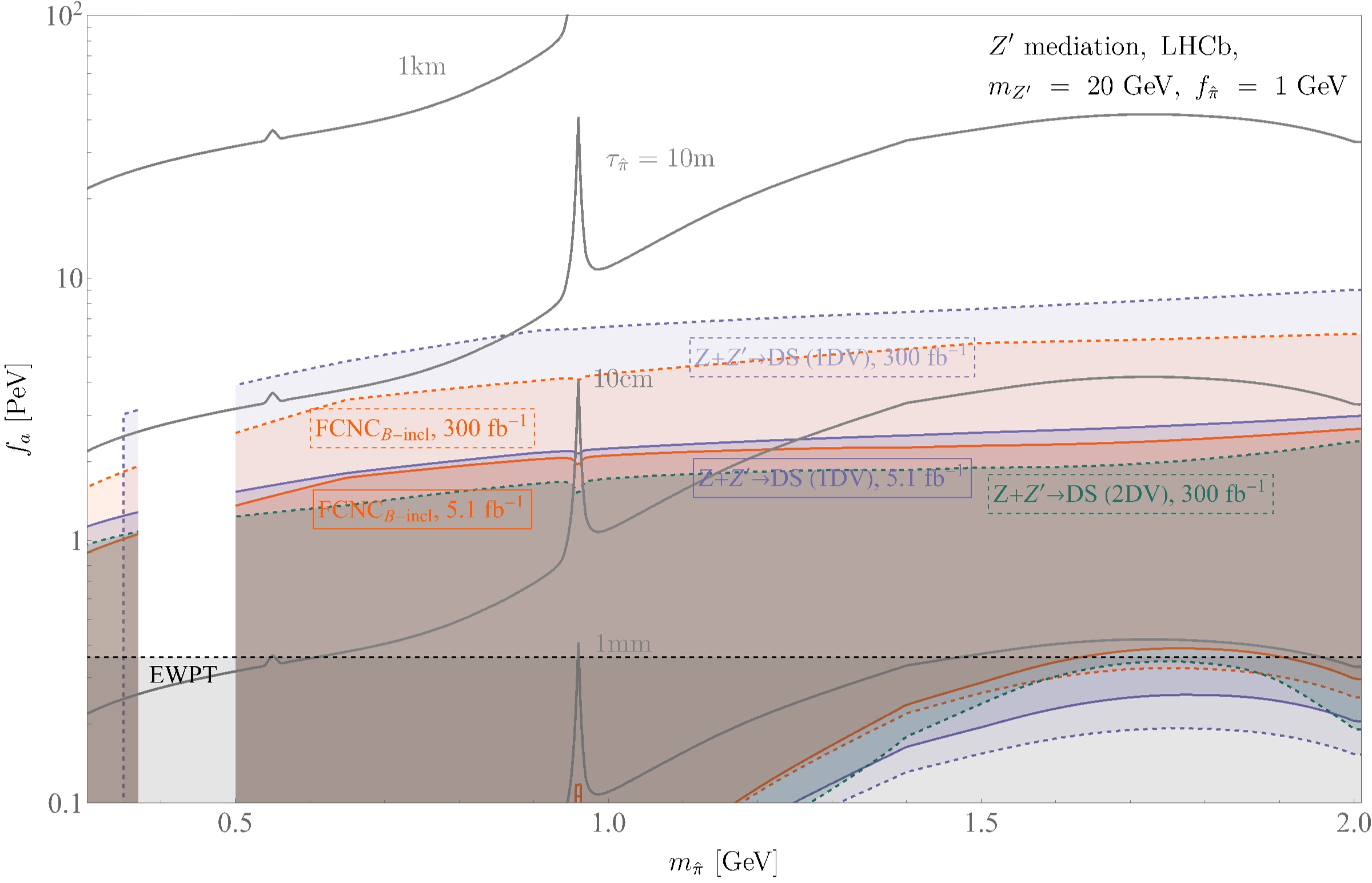}
\caption{Similar to Fig.~\ref{fig:modeldependent_1} but for LHCb limits.}
\label{fig:modeldependent_2}
\end{figure}

\noindent \textbf{Constraints from LHCb} With the same benchmarks, we plot the current and projected future LHCb limits on such dark shower signals in Fig.~\ref{fig:modeldependent_2}. In the dark $Z'$ model, with $m_{Z^\prime}=20$~GeV,  the dark shower signals from the light $Z^\prime$ dominate the exclusion limit, similar to the CMS case. As discussed in Sec.~\ref{sec:DarkShower}, LHCb has the advantage of a lower muon $p_T$ threshold and the coverage of high $|\eta|$ region for a light $Z'$. It is thus relatively advantageous compared to the CMS in the dark shower searches with a light dark $Z'$. Similarly, LHCb has better reaches in FCNC decays due to the unsuppressed $B$ meson production in the forward region and the conservative approach in our recasting of the CMS scouting search. For the heavy fermion mixing model, LHCb and CMS have similar reaches in the dark shower signals. 

 \begin{figure}[h!]
\centering
\includegraphics[width=7.5 cm]{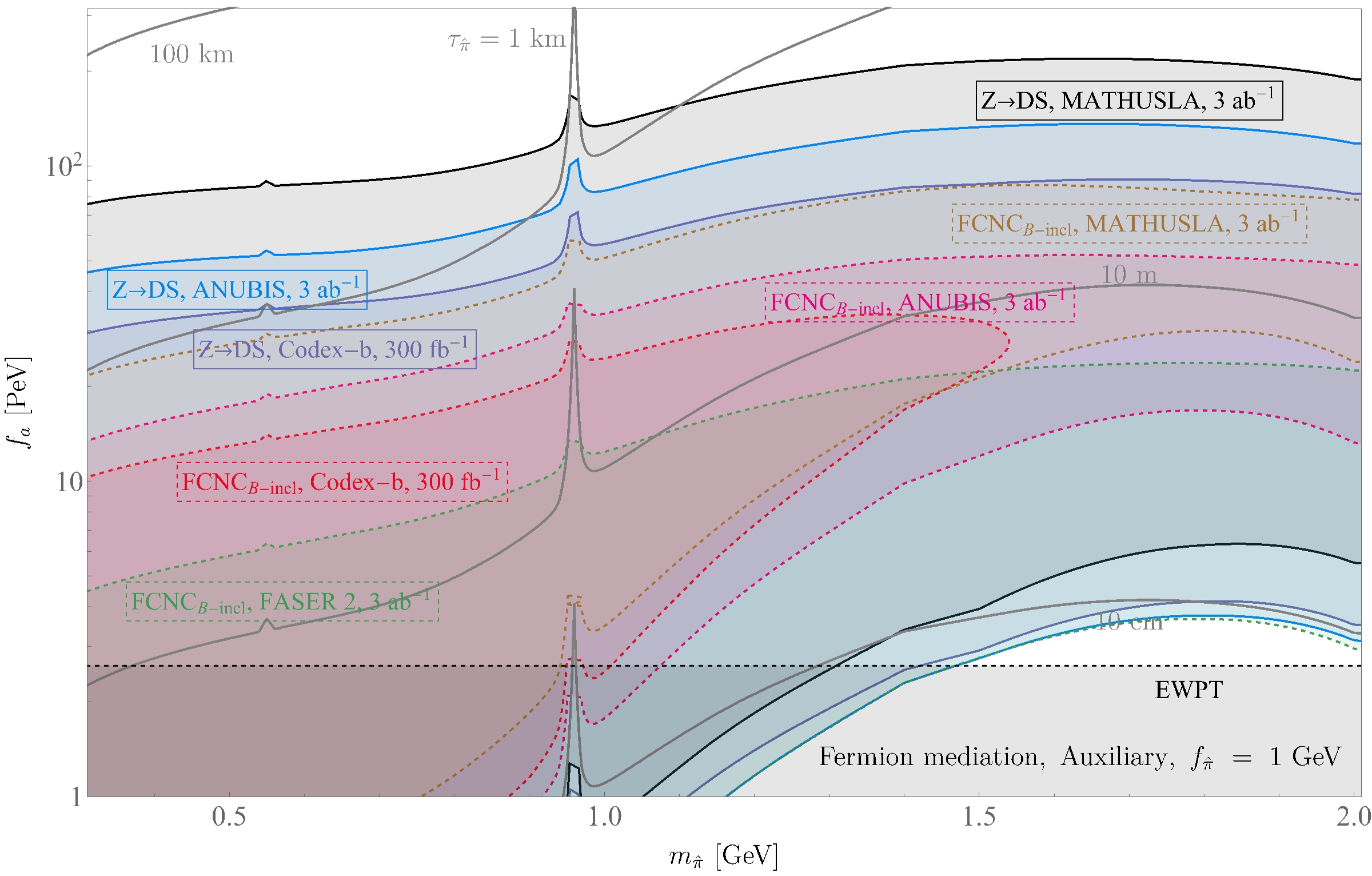}
\includegraphics[width=7.5 cm]{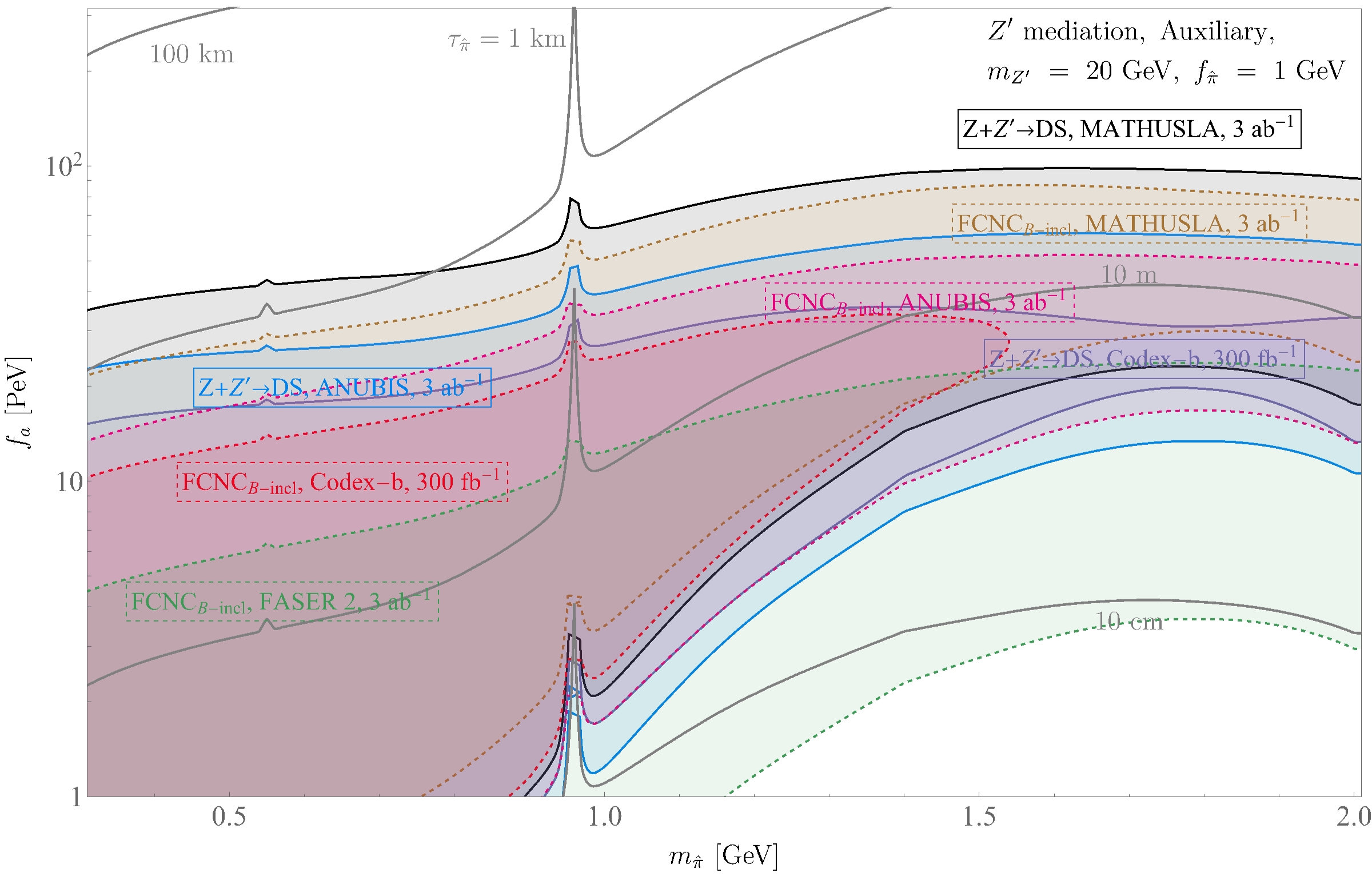}
\caption{Similar to Fig.~\ref{fig:modeldependent_1} but for auxiliary detector limits. For MATHUSLA we adopt the weaker updated design limits.}
\label{fig:modeldependent_3}
\end{figure}

\noindent\textbf{Constraints from Auxiliary Detectors} Following a similar procedure, the model-specific limits for both benchmark models at (HL-)LHC auxiliary detectors are plotted in Fig.~\ref{fig:modeldependent_3}. The dark shower signals are accessible at MATHUSLA, ANUBIS, and Codex-b. As they are sensitive to dark pions with $c\tau(\hat{\pi}) \gtrsim 10$~m with low backgrounds, the reach of $f_a$ can get close to or even exceed $10^2$~PeV. Meanwhile, all auxiliary detectors considered can probe the FCNC decay signals, though for $f_{\hat{\pi}}=1$~GeV, they are weaker than the dark shower reaches. Nevertheless, they can still reach $f_a$ values comparable to the dark shower limits in other detectors thanks to the detectors' reach in the long-lifetime regime.

\begin{figure}[h!]
\centering
\includegraphics[width=7.5 cm]{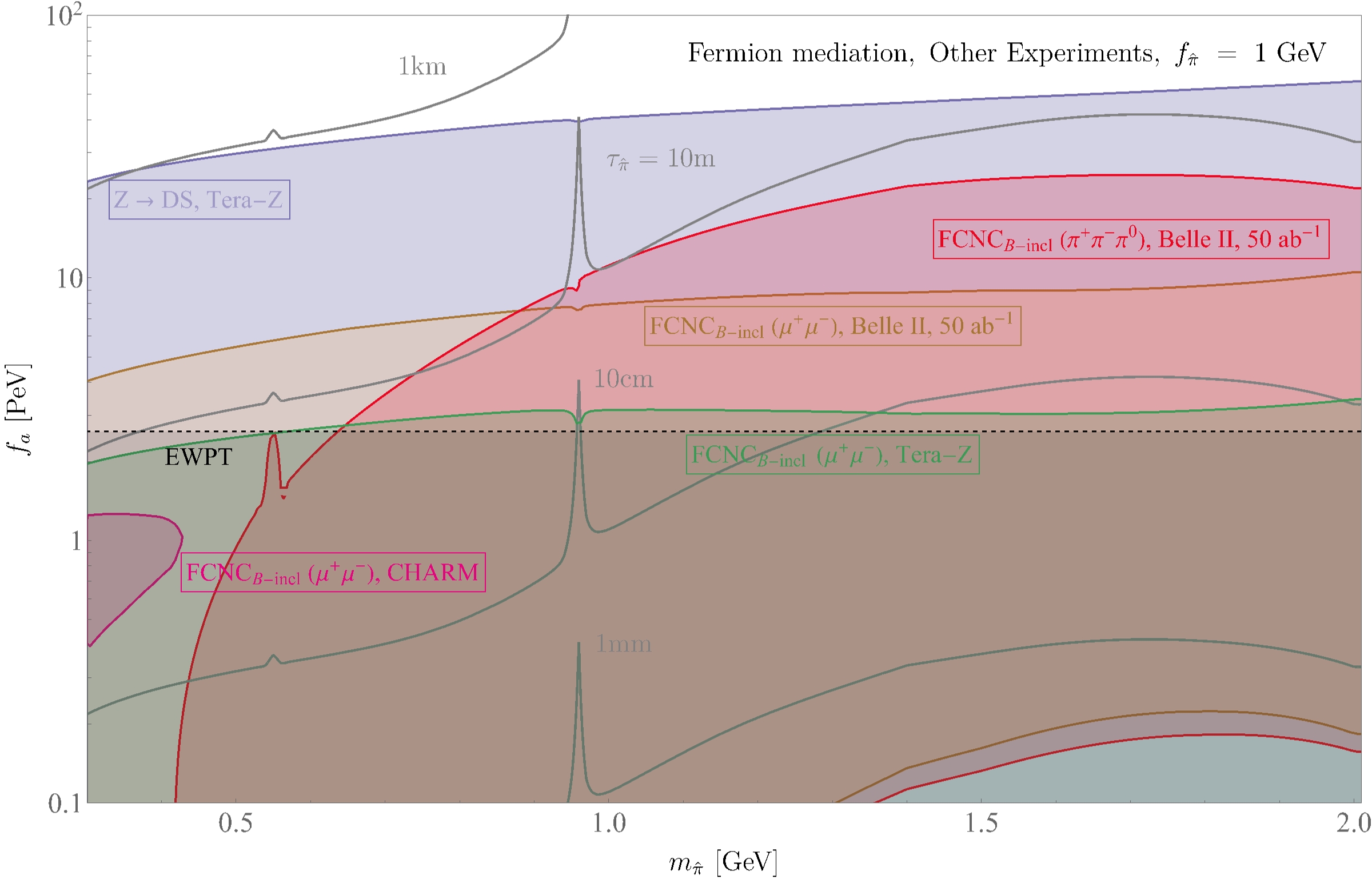}
\includegraphics[width=7.5 cm]{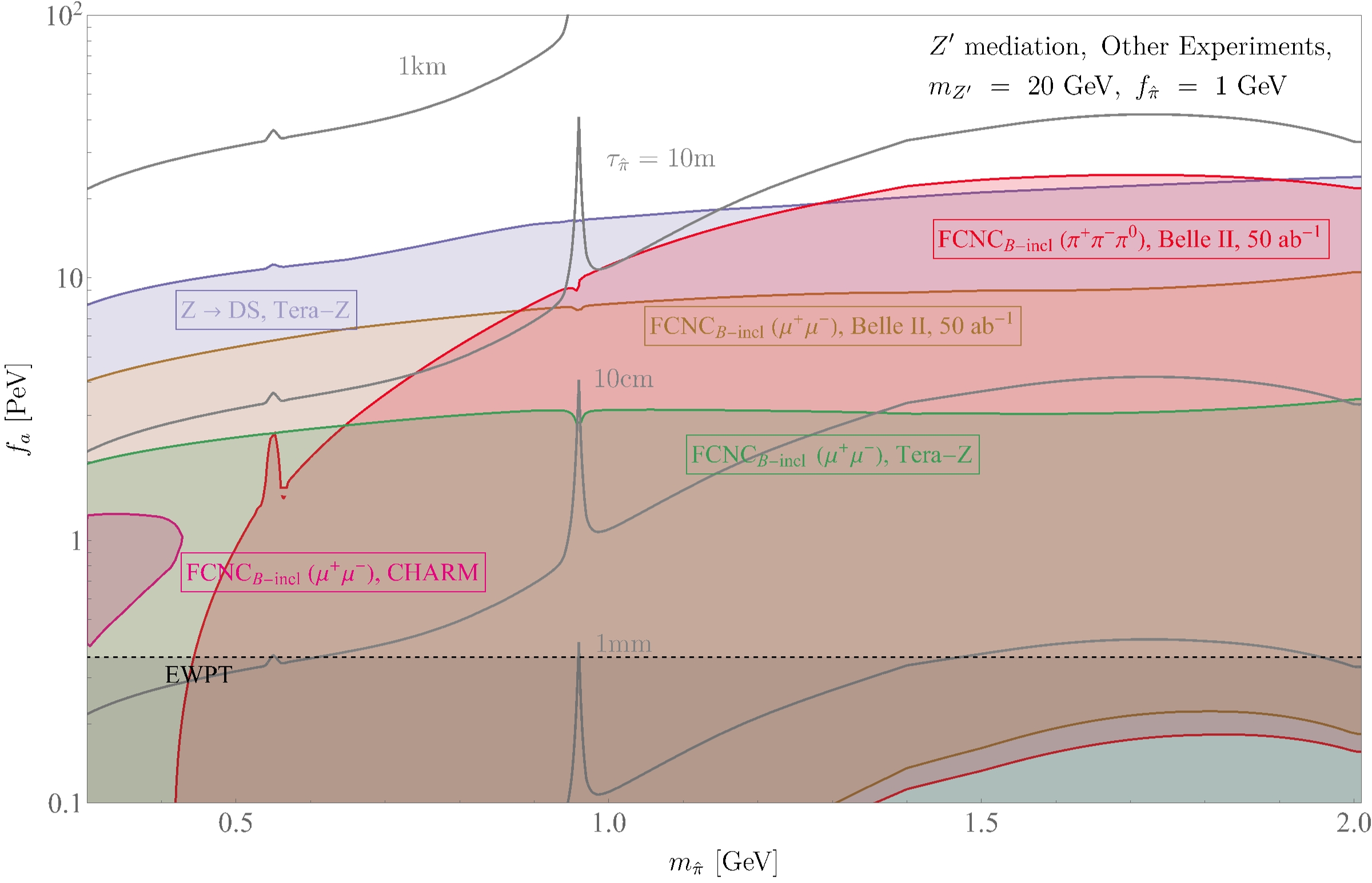}
\caption{Similar to Fig.~\ref{fig:modeldependent_1} but for intensity frontier limits.}
\label{fig:modeldependent_4}
\end{figure}

\noindent\textbf{Constraints from Rare and Precision Frontiers}
In Fig.~\ref{fig:modeldependent_4} we list constraints from several other experiments. The dark pions from FCNC $B$ decays can be tested at Belle II and fixed target experiments. Together, they cover a wide range of parameter space, led by the Belle II constraints. Note that $\pi^+\pi^-\pi^0$ and $\mu^+\mu^-$ decay modes at Belle II give comparable limits. The reach of the  $\pi^+\pi^-\pi^0$ mode is slightly more favorable for $m_{\hat{\pi}}>1.2$~GeV thanks to its large branching ratio. We also show the $Z$ initiated dark shower limits at Tera-$Z$, which also reach parameter regions far beyond what CMS and LHCb can do at the HL-LHC.

\section{Summary and Discussion \label{sec:Summary}}
In this work, we present a general study of the dark shower phenomenology of the EW portals. Specifically, we consider dark showers initiated from the decays of the SM $Z$, $h$, and possible $Z'$ and scalar $\phi$ bosons near the EW scale. If these particles are heavy enough compared to the dark QCD scale, their decays into the dark QCD sector will produce multiple dark hadrons. Unlike other high-scale mediation models, the total energies of the dark showers are controlled by the masses of these EW scale bosons, and in most events, no hard objects are produced to help triggering these types of dark shower events. These features make general search methods at colliders ineffective for these events. Specialized search strategies, such as data scouting and data parking at CMS, which significantly lower the muon $p_T$ thresholds, are suited for this task, utilizing the fact that light dark hadrons are typically LLPs with macroscopic decay lengths. The LHCb experiment, with its excellent vertex resolution, is also capable of searching for these dark shower events, especially for relatively short decay lengths. For longer decay lengths, auxiliary detectors far from the collision points are advantageous for detection. New experimental techniques and algorithms such as Refs.~\cite{CMS:2024bvl,Gorkavenko:2023nbk} may allow detectors like CMS or LHCb to probe longer lifetimes. Their current reach on proposed EW-portal LLP signals is likely to be limited, and right now there is not sufficient information for a good estimation for future projections.

The details of the dark showers produced from the EW portals mostly depend only on the parameters of the dark hadrons and dark QCD, but not the mediators. This allows us to study their collider reaches in a simplified approach. The rate of producing observable displaced (dimuon) events is governed by the exotic decay branching ratios of $Z$, $h$ into dark quarks, or the mixing parameters between the dark $Z'$, $\phi$ with $Z$, $h$ (and the branching ratio of dark pion decaying into dimuon). The experimental reaches of the rate depend on the dark pion masses and their lifetimes. For each experiment, we can project the reach limits on the production rate vs.\/ decay length plane for a given dark pions mass. These limits can be easily applied to obtain constraints on model parameters for any specific models that produce EW scale dark showers. Consequently, it is a valuable exercise to provide such limits. In addition to obtaining the current limits by recasting the current searches, we also make projections on future reaches for CMS scouting searches and LHCb, as well as other proposed future detectors or experiments, based on some conservative assumptions. The projections could be further enhanced if more improvements in experimental searches can be realized or better strategies can be employed. For example, the data parking approach retains the full event information and could be even more powerful than data scouting when searching for dark shower events. However, there is no current search available for us to make realistic projections.

The light dark hadrons can also be produced in FCNC $B$ decays with similar search strategies as the dark shower searches. The experimental reaches can also be studied. The bounds are put on the branching ratio of FCNC $B$ decay into dark pions vs. dark pion lifetime plane for a given dark pion mass. In addition to LHC experiments discussed in the dark shower search, the FCNC decays can also be searched at fixed-target or beam-dump experiments, the $B$-factory Belle II, and the LHC forward experiments such as the FASER. The projections of their reaches are also obtained. The branching ratio of the FCNC meson decay into a dark pion mostly just depends on the effective ALP decay constant $f_a$, plus a mild logarithmic dependence of the UV cutoff of the underlying theory. It provides independent information from the dark shower events, which depend on their productions from the EW bosons, and therefore serves as a complementary test.

The relative strengths of various experiments are model dependent. To compare them, specific models are required. The bounds can then be translated into bounds on the model parameters. We illustrate this with two benchmark models with dark showers generated dominantly from $Z$ and $Z'$. Reach limits from different experiments can be cast in the same parameter space for comparison. We found that most experiments can probe regions beyond the current bounds allowed by the EWPT. For $f_{\hat{\pi}}= 1$~GeV, the reaches of the dark shower searches are typically somewhat stronger than the FCNC decay searches, although for larger $f_{\hat{\pi}}$ the trend could be reversed. CMS scouting search and LHCb generally give comparable limits, with LHCb winning in the case of a light $Z'$. For longer decay lengths, the auxiliary detectors, in particular, MATHUSLA, will have the most powerful reach thanks to its location and large volume. Of course, these two benchmark models only represent a small set of the large possible model space, and the comparison of different experiments within these models should not be taken as general conclusions. Nevertheless, they demonstrate how the results obtained in this paper can be applied to specific models.

\acknowledgments

We thank Ennio Salvioni for collaboration in the early stage of this work. We also thank Cristiano Alpigiani, Reuven Balkin, Elias Bernreuther, Matthew Citron, Quentin Bonnerfoy, Johnathan Feng, Simon Knapen, Suchita Kulkarni, Zhen Liu, Jingyu Luo, Navin McGinnis, Jessie Shelton, Mattew Strassler, Indaria Suarez, Yuhsin Tsai, Mike Williams, Keping Xie for useful discussions. H.C. was supported by the Department of Energy under grant DE-SC0009999. X.J. was supported in part by the National Natural Science Foundation of China under grant 12342502.  L.L. is supported by DOE grant DE-SC0010010. This work was performed in part at Aspen Center for Physics, which is supported by National Science Foundation grant PHY-2210452.

\appendix

 \section{Dimuon Vertex Trigger Efficiencies at HL-LHC}
 \label{app:Trigger}
\begin{figure}[h!]
\centering
\includegraphics[width=6 cm]{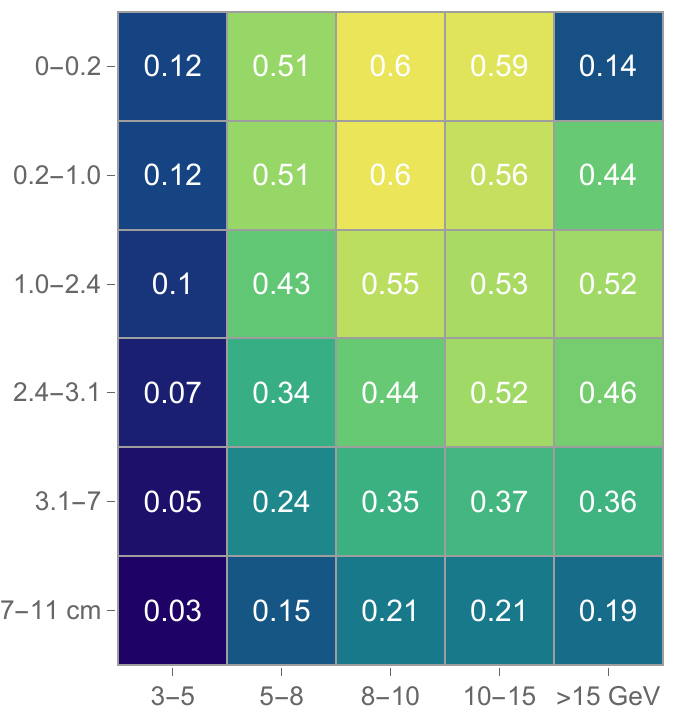}
\includegraphics[width=6 cm]{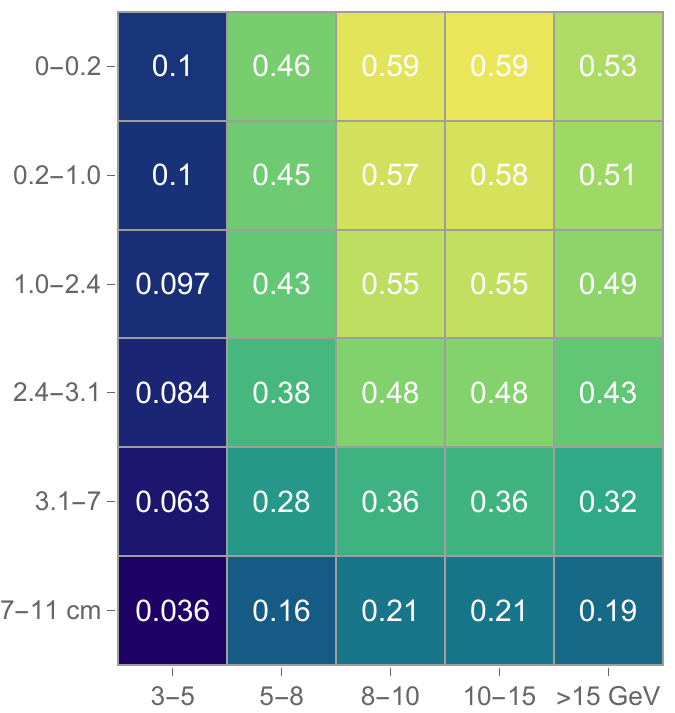}
\caption{DV trigger efficiency for different $p_{T,\mu 2}$ and $l_{xy}$. \textbf{LEFT: }The efficiency taken from the CMS reference~\cite{CMS:2021ogd}, estimated from samples of $B\to \mu^+\mu^- +X$ inclusive decays assuming a long-lived intermediate particle lighter than 3~GeV. \textbf{RIGHT: }The approximated efficiency assuming its is the product of two independent functions.\label{fig:tableLHC}
}
\end{figure}

 To extend the CMS efficiency of dimuon DVs~\cite{CMS:2021ogd}, we first make the assumption that of each dimuon DV's trigger efficiency can be approximated by the products of two functions. One only depends on the minimum transverse momentum of the two muons $p_{T,\mu 2}$ and the other depends only on the DV's transverse displacement $l_{xy}$. The functions can be determined by matching with the efficiency table provided by the supplement figures of Ref.~\cite{CMS:2021ogd}. The comparison between the original efficiency and the approximation is shown in Fig.~\ref{fig:tableLHC}. In particular, we found the dependence on $p_{T,\mu 2}$ is similar to the DV selection efficiency when we require $p_{T,\mu 1(2)} > 5(3)$~GeV and $|\eta_{\mu 1,2}|<2.4$.
 
\begin{figure}[h!]
\centering
\includegraphics[width=10 cm]{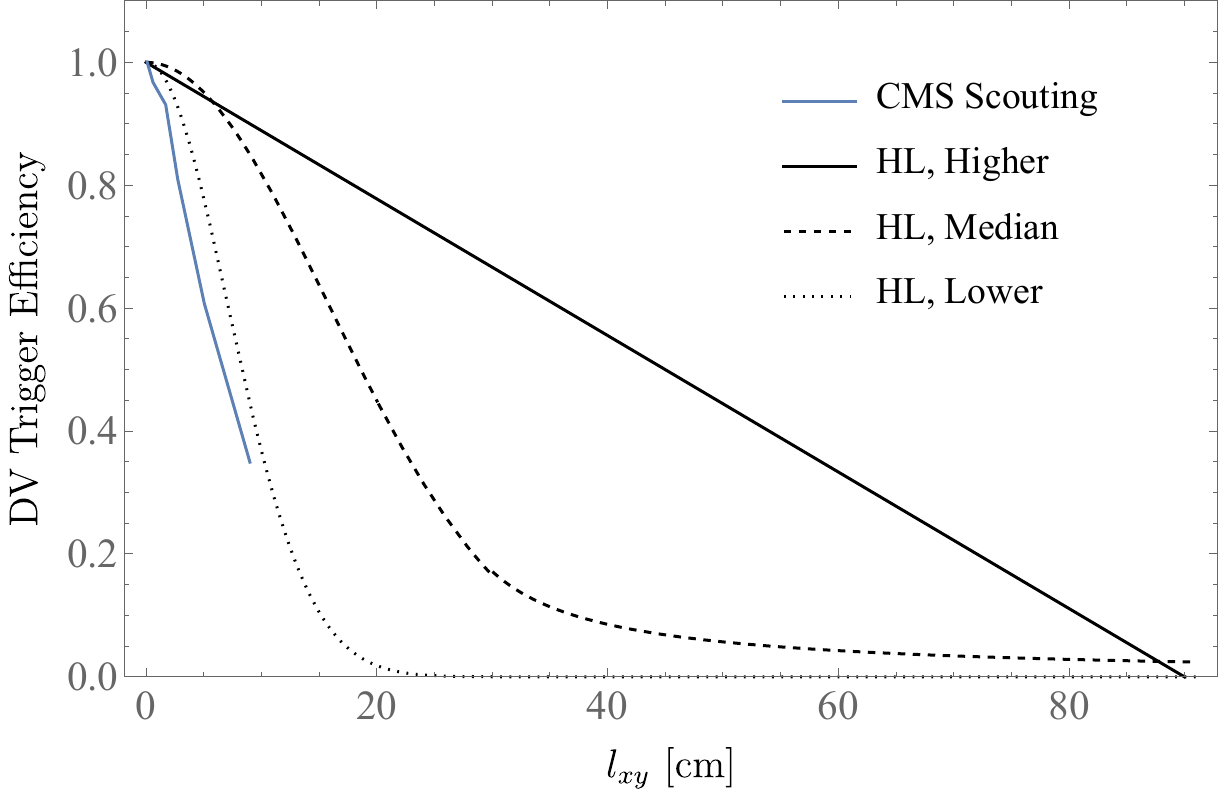}
\caption{The efficiency of DV and individual lepton tracks as a function of $l_{xy}$. Black curves are three different scenarios.}
\label{fig:LargeIPTracks}
\end{figure}

\begin{figure}[h!]
\centering
\includegraphics[width=4.7 cm]{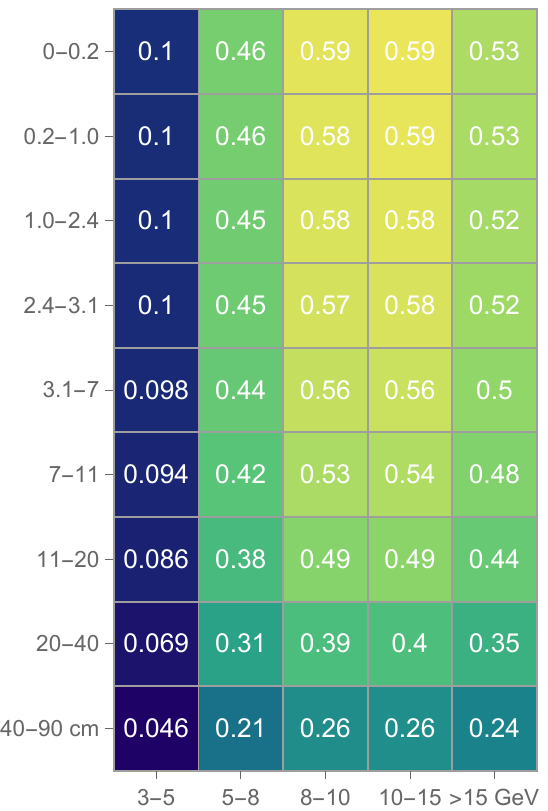}
\includegraphics[width=4.7 cm]{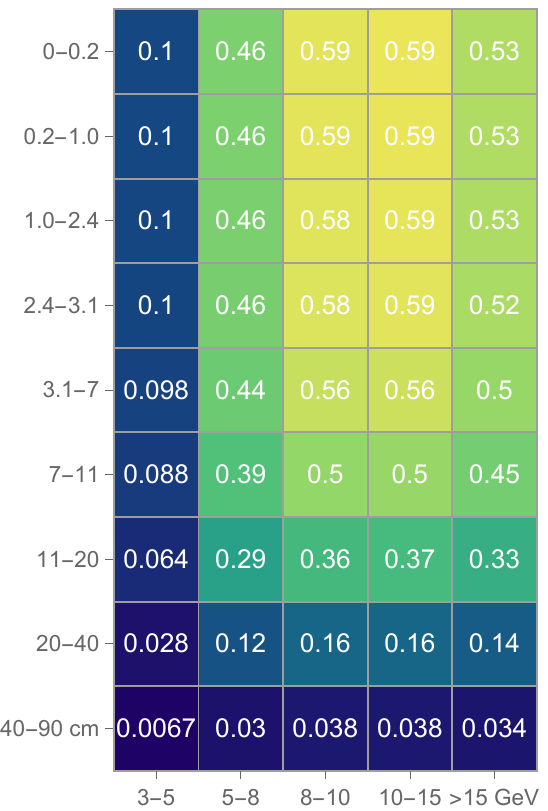}
\includegraphics[width=4.7 cm]{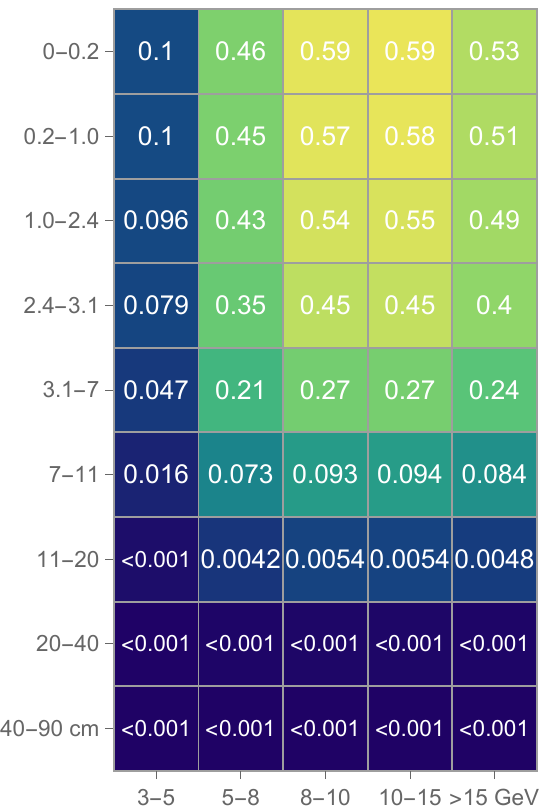}
\caption{Three binned DV trigger efficiencies from different HL-LHC scenarios, assuming they are the product of $p_T$ and $l_{xy} $ dependent factors. The overall efficiency decreases from left to right.}
\label{fig:HLscenaros}
\end{figure}

To project the known result to the high-luminosity era, the trigger efficiency's dependence on the transverse displacement $l_{xy}$ needs to be revamped, especially for $l_{xy}>11$~cm cases which are not covered by current studies. To estimate, we consider the performance of the novel tracking techniques for large impact parameter tracks, such as Ref.~\cite{ATLAS:2017zsd}. In Fig.~\ref{fig:LargeIPTracks} we show the conjectured $l_{xy}$ dependences of trigger efficiency in the HL-LHC era, with three scenarios to contain various possibilities. Here the ``lower" scenario serves as a conservative extension of the current CMS performance beyond $l_{xy}\in[0,11]$~cm. In contrast, the ``higher" scenario assumes a linear $l_{xy}$ dependence, which significantly improves the signal reach when $l_{xy}>11$~cm. The ``median" scenario is inspired by the performance gain with the large IP track algorithm~\cite{ATLAS:2017zsd}. For comparison, the extracted CMS trigger efficiency's $l_{xy}$ dependence corresponding to the right panel of Fig.~\ref{fig:tableLHC} is also shown. The trigger efficiency values of each $p_T$ and $l_{xy}$ bin are shown in Fig.~\ref{fig:HLscenaros}.

\section{Benchmarks for Model-Specific Projections}
\label{app:benchmark}
\subsection{Heavy Doublet Fermion Mixing Model}
We consider the ``Scenario 1'' in Ref.~~\cite{Cheng:2021kjg} where $\byukt=0$. We assume that the heavy dark quark masses are degenerate, $\bm{M} = M \bm{1}_2$, the light dark quark mass matrix $\bm{\omega}$ is diagonal, and $CP$ is conserved. Because $\byukt=0$, the corrections to the light quark masses vanish at leading order. 
The effective dark pion decay constants are
\begin{equation}
\frac{1}{f_a^{(b)}} = \frac{f_{\hat{\pi}}}{4} \mathrm{Tr} \big[ \sigma_b \bm{Y}^\dagger \bm{M}^{-2} \bm{Y} \big] = \frac{f_{\hat{\pi}}}{4M^2} \mathrm{Tr} \big[ \sigma_b \bm{Y}^\dagger \bm{Y} \big]\,,
\end{equation}
with
\begin{equation}
\mathrm{Tr} \big[ \sigma_1 \bm{Y}^\dagger \bm{Y} \big] = 2 (y_{11} y_{12} + y_{21} y_{22} )\,,\qquad \mathrm{Tr} \big[ \sigma_3 \bm{Y}^\dagger \bm{Y} \big] = y_{11}^2 - y_{12}^2 + y_{21}^2 - y_{22}^2\,~,
\end{equation}
They can be further simplified by setting for example $y_{21} = y_{22} = 0$, then
\begin{equation}
\frac{1}{f_a^{(1)}} = \frac{f_{\hat{\pi}}}{2M^2} \, y_{11} y_{12}\,, \qquad \frac{1}{f_a^{(3)}} = \frac{f_{\hat{\pi}}}{4M^2}  ( y_{11}^2 - y_{12}^2 )\,.
\end{equation}
If we choose $y_{11}/y_{12} = 1 + \sqrt{2}\,$, then $f_a^{(1)}=f_a^{(3)} = f_{\hat{\pi}} y^2 / [ 2(1 + \sqrt{2})M^2 ]$, where $y \equiv y_{11}\,$. Introducing a small $\widetilde{\bm{Y}} \ll \bm{Y}$ makes $\hat{\pi}_2$ decay with lifetime much longer than those of the other two dark pions, without otherwise altering the above picture.

To evaluate the constraints from EWPT on this benchmark model, in principle we need to calculate both $S$ and $T$. We argue, however, that $S$ is parametrically suppressed and can be neglected. This is best illustrated in a toy one-flavor model with $\widetilde{Y} = 0$. Applying the general results of Ref.~\cite{Albergaria:2023nby} (see also Refs.~\cite{Lavoura:1992np,Bizot:2015zaa}) we find
\begin{equation}
S \simeq  \frac{2 N_d}{45 \pi M^2} \Big( \frac{5 Y^2 v^2}{2} - 3 m_Z^2 (1 - 2 s_W^2 +  2 s_W^4 ) \Big) \,, \qquad T \simeq \frac{N_d Y^4 v^4}{48 \pi s_W^2 m_W^2 M^2}\,,
\end{equation}
leading to
\begin{equation}
\widehat{S} / \widehat{T} \simeq \frac{g^2}{3 Y^2} \approx \frac{1}{7 Y^2}\,,
\end{equation}
which is valid for $Y \gtrsim 1$. Since $\widehat{S}$ and $\widehat{T}$ are constrained at a similar level, as long as the dark Yukawa coupling has size $Y \gtrsim 1$ it is justified to derive EWPT constraints by retaining only $T$. Back to our benchmark model, we find
\begin{equation}
T \simeq \frac{ N_d \,y_{11}^4 v^4  }{48 \pi s_W^2  m_W^2 M^2  } \Big( 1 + \frac{y_{12}^2}{y_{11}^2} \Big)^2 \approx \frac{ 1.4\, N_d \,y^4 v^4  }{48 \pi s_W^2  m_W^2 M^2  } \, ,
\end{equation}
where in the last step we have assumed $y_{11}/y_{12} = 1 + \sqrt{2}$. Given the experimental determination $T = 0.03 \pm 0.12$~\cite{ParticleDataGroup:2022pth}, we set our approximate $95\%$~CL constraint by requiring $T < 0.24$.

\subsection{$Z-$Dark $Z^\prime$ Mixing Model}
The dark $Z'$ acquires its mass from a scalar VEV $\langle \Phi \rangle = v_\Phi$ which spontaneously breaks the dark $U(1)'$ gauge group. The light dark quarks can obtain mass contributions from couplings to $\Phi$, in addition to the intrinsic ones,
\begin{equation}
 \sum_{i,\,j \,=\, 1}^N \Big( \overline{\psi}_{L i} m_{ij}  \psi_{R j} +  \overline{\psi}_{L i} \zeta^1_{ij} \psi_{Rj} \Phi + \overline{\psi}_{R i} \zeta^2_{ij} \psi_{Lj} \Phi  + \mathrm{h.c.} \Big) \, .
 \end{equation}
We assume that the two flavors of the light dark quarks have vector-like charges $x_1, x_2$ under $U(1)'$, with $x_1-x_2=x_\Phi$ so the off-diagonal masses can be generated from the Yukawa terms. The dark quark mass matrix then takes the form
\begin{equation}
\bmpsi = \begin{pmatrix} m_1 & y_1 v_\Phi \\ y_2 v_\Phi  & m_2 \end{pmatrix}\, .
\end{equation}
A motivated and simple limit is $y_2 \to 0$, which leads to $CP$ conservation (hence $\mathrm{Tr}(\sigma_2 \bm{X}^\prime_A) = \mathrm{Tr}(\sigma_2 \bm{X}^\prime_V) = \mathrm{Tr}[i\sigma_{1,3}(\bm{\zeta}^\prime - \bm{\zeta}^{\prime\,\dagger})] = 0$) and also has the advantage that compact analytical expressions can be derived for the relevant traces. In addition, we assume the further simplification $m_2 \to 0\,$: in this case one dark quark is massless, $\mathrm{Tr} (\bm{m}_{\psi^\prime}) = (m_1^2 + y_1^2 v_\Phi^2)^{1/2}$ and we find
\begin{align}
\mathrm{Tr}&(\sigma_1 \bm{X}^\prime_{A,V}) = -  \frac{ m_1 y_1 v_\Phi  }{m_1^2 + y_1^2 v_\Phi^2} (x_1 - x_2) \,,\qquad \mathrm{Tr}(\sigma_3 \bm{X}^\prime_{A,V}) =  \frac{ \{ -\,y_1^2 v_\Phi^2, m_1^2 \}}{ m_1^2 + y_1^2 v_\Phi^2   }(x_1 - x_2)\,, \;  \nonumber \\  
&\mathrm{Tr}[i\sigma_2 (\bm{\zeta}^\prime - \bm{\zeta}^{\prime\,\dagger})] =  - 2 y_1   \frac{m_1}{ ( m_1^2 + y_1^2 v_\Phi^2  )^{1/2} } \,, \qquad \mathrm{Tr}(\sigma_0\bm{X}^\prime_V) = x_1 + x_2\,. \label{eq:m2_0}
\end{align}
We choose the benchmark of $y_1 v_\Phi=m_1$ so that both $CP$-odd dark pions $\hat{\pi}_{1,3}$ decay with the same lifetime. With the charge assignment $x_1 = - 1 = - x_2$, resulting in $\mathrm{Tr}(\sigma_{1,3} \bm{X}^\prime_A ) = 1$. For definiteness, we also set the $U(1)'$ gauge coupling to $g_D = 0.25$ similar to the electromagnetic coupling in the SM.

\bibliographystyle{JHEP}
\bibliography{darkshower}

\end{document}